\def\jobname{Constraining-SFDM-AntliaB}
\documentclass{aa} 

\usepackage{graphicx}
\usepackage{txfonts}

\usepackage{float}
\usepackage{array}
\usepackage{hyperref}
\usepackage{physics}
\usepackage{amsfonts}
\usepackage{amsmath}
\usepackage{amssymb}
\usepackage{xcolor}
\usepackage{placeins}
\usepackage{multirow}

\DeclareMathOperator{\sinc}{sinc}
\newcommand{\hi}{\ensuremath{\mathrm{H\,\textsc{i}}}}

\hypersetup{pdftitle={Constraining Scalar Field Dark Matter with Antlia B}}

\begin{document} 

   \title{The MUSE-Faint survey}
   \subtitle{III. Constraining Scalar Field Dark Matter with Antlia B}

   \author{Mariana P. Júlio \inst{1, 2, 3} \and Jarle Brinchmann \inst{1, 2, 4} \and Sebastiaan L. Zoutendijk \inst{4} \and Justin I. Read \inst{5} \and Daniel Vaz \inst{1, 2} \and Sebastian Kamann \inst{6} \and Davor Krajnović \inst{3} \and Leindert A. Boogaard \inst{7} \and Matthias Steinmetz\inst{3} \and Nicolas Bouché\inst{8}}

   \institute{Instituto de Astrofísica e Ciências do Espaço, Universidade do Porto, CAUP, Rua das Estrelas, PT4150-762 Porto, Portugal \and Departamento de Física e Astronomia, Faculdade de Ciências, Universidade do Porto, Rua do Campo Alegre 687, PT4169-007 Porto, Portugal \and Leibniz-Institut für Astrophysik Potsdam (AIP), An der Sternwarte 16, D-14482 Potsdam, Germany \and Leiden Observatory, Leiden University, P.O. Box 9513, 2300 RA Leiden, The Netherlands \and University of Surrey, Physics Department, Guildford, GU2 7XH, UK \and Astrophysics Research Institute, Liverpool John Moores University, IC2 Liverpool Science Park, 146 Brownlow Hill, Liverpool L35RF, United Kingdom \and Max Planck Institute for Astronomy, K\"onigstuhl 17, 69117 Heidelberg, Germany \and Univ. Lyon, Univ. Lyon1, ENS de Lyon, CNRS, Centre de Recherche Astrophysique de Lyon UMR5574, 69230, Saint-Genis-Laval, France}
             
   \date{Received 19 May 2023 / Accepted 21 July 2023}

    \abstract
    {}
     {We use the stellar line-of-sight velocities of Antlia B (Ant B), a faint dwarf galaxy in the NGC 3109 association, to derive constraints on the fundamental properties of scalar field dark matter (SFDM), which was originally proposed to solve the small-scale problems faced by cold dark matter models.}
     {We used the first spectroscopic observations of Ant B, a distant (d $\sim$ 1.35 Mpc) faint dwarf (M$_V = -9.7$, M$_\star \sim 8\times10^5$M$_\odot$), from MUSE-Faint, a survey of ultra-faint dwarfs conducted using the Multi Unit Spectroscopic Explorer. By measuring the line-of-sight velocities of stars in the $1'\times 1'$ field of view, we identified 127 stars as members of Ant B, which enabled us to model its dark matter density profile with the Jeans modelling code \textsc{GravSphere}. We implemented a model for SFDM into \textsc{GravSphere} and used this to place constraints on the self-coupling strength of this model.}
     {We find a virial mass of ${M_{200} \approx 1.66^{+2.51}_{-0.92}\times 10^9}$ M$_\odot$ and a concentration parameter of ${c_{200}\approx 17.38^{+6.06}_{-4.20}}$ for Ant B. These results are consistent with the mass-concentration relations in the literature. We constrain the characteristic length scale of the repulsive self-interaction $R_{\text{TF}}$ of the SFDM model to $R_{\text{TF}} \lesssim 180$ pc ($68\%$ confidence level), which translates to a self-coupling strength of $\frac{g}{m^2c^4}\lesssim 5.2 \times 10^{-20}$ eV$^{-1}$cm$^3$. The constraint on the characteristic length scale of the repulsive self-interaction is inconsistent with the value required to match observations of the cores of dwarf galaxies in the Local Group, suggesting that the cored density profiles of those galaxies are not caused by SFDM.}{}

   
    \titlerunning{The MUSE-Faint Survey. III.}
	\authorrunning{M. P. Júlio et al.}
	
	\maketitle


\section{Introduction}\label{introduction}
\vspace{-1mm}
Since dark matter was first proposed as an additional component to explain the masses of galaxy clusters \citep{zwicky37} and the flat observed rotation curves of galaxies (\citealt{rubin}, \citealt{bosma1981}), many more indications for the existence of dark matter have been found (see \citealt{Bertone_2018} for an extensive review of the history of dark matter). The presence of this non-baryonic matter is required to explain the behaviour of most structures existing in the Universe and the vast majority of cosmological observations favour a cold-dark-matter-plus-dark-energy scenario (the $\Lambda$CDM model), from sub-galactic and galactic scales (dwarf galaxies, e.g. \citealt{read2019}), to the scale of galaxy clusters (e.g. \citealt{massey2018} and large-scale structures (e.g.  \citealt{lss_review}, \citealt{Baur_2016}), all the way up to cosmological scales (anisotropies of the cosmic microwave background, \citealt{planck}). The $\Lambda$CDM model is even able to explain the offsets between mass and light in weak lensing systems (e.g. \citealt{Harvey_2015}; see e.g. \citealt{ferreira_2021} for a review).  \par
$\Lambda$CDM has been very successful at predicting and explaining the large-scale structure of the Universe and its evolution with time (e.g. \citealt{cosmsim_vogel14} and \citealt{cosmsim_joop15}), which has led to it now being considered the standard model of cosmology. However, when we look at smaller scales, with length scales smaller than $\sim 1$ Mpc and mass scales smaller than $\sim 10^{11}$ M$_\odot$, pure dark matter structure formation simulations in the standard model starts to face numerous challenges \citep{bullock}. Some of the well-known problems associated with small scales are: the missing satellites problem (galaxies like the Milky Way should have significantly more bound dark matter subhaloes than the number of observed satellite galaxies; \citealt{klypin99} and \citealt{Moore_1999}); the too-big-to-fail (TBTF) problem (dwarf galaxies are expected to be hosted by haloes that are significantly more massive than is indicated by the measured galactic velocity; \citealt{tbtf11}); and the core-cusp problem (the observed cores of many dark-matter-dominated galaxies are both less dense and less cuspy than predicted by the pure dark matter $N$-body simulations in the standard model; \citealt{Flores_1994} and \citealt{1994moore})\footnote{Note that the TBTF problem can also be understood as the core-cusp problem but for satellites. That is to say, the TBTF can be solved if (some) satellite dwarf galaxies have dark matter cores similar to those found in isolated dwarfs (see e.g. \citealt{read2006}, \citealt{read_erkal19}).}. One way to resolve these discrepancies is to posit that we need a modified gravity model (see e.g. \citealt{monds} for a review). While these offer an interesting pathway, they face several observational challenges at all scales (e.g. \citealt{cluster2006}, \citealt{zhao2008}, \citealt{dodelson}, \citealt{globcluster}, and \citealt{read2019}). In this paper, we therefore choose to focus on ways to resolve these small-scale challenges within the $\Lambda$CDM model and its variants.\par 
Three major scenarios have been proposed to solve the small-scale problems in $\Lambda$CDM. The first one implies that there might be problems with the data from the observations due to poor resolution (e.g. \citealt{blok}) or misinterpretation of the data (e.g. \citealt{valenzuela2007} and \citealt{oman2015}). Another possibility is the lack of modelling of baryonic physics in the simulations. Adding baryons to the simulations can straightforwardly solve the missing satellite problem by an appropriate mapping of visible galaxies to dark subhaloes (e.g. \citealt{read_erkal19}). However, the core-cusp and TBTF problems are harder as they seem to imply that baryons need to physically move dark matter out of the centres of galaxies. Mechanisms have been proposed to address this, whereby repeated gas inflow-outflow and/or dynamical friction from dense clumps of gas or stars cause the gravitational potential to fluctuate over time. This enables energy to be transferred to dark matter particle orbits, pushing dark matter out of galaxy centres and making them more cored (e.g. \citealt{navarro1996}, \citealt{zant}, \citealt{nipoti}, \citealt{read2019}). The last class of solutions, and perhaps the most exciting, conjectures that the nature of dark matter itself is different from the current paradigm. The proposed alternatives to CDM cover a wide range of particle masses, from macroscopic objects (e.g. massive compact halo objects; \citealt{machos}, and primordial black holes; \citealt{pbhs}) to axion-like particles \citep{preskill83}, as well as a wide range of different interactions between the dark matter particles, such as self-interaction (SI) through scattering \citep{firstsidm} or annihilation \citep{kaplinghat}. Recently, DM models that involve ultra-light particles that have masses low enough to exhibit wave-like behaviour on astrophysical scales have been emerging. The free-field case of these models is called fuzzy dark matter (FDM; \citealt{dine83}, \citealt{preskill83}, \citealt{firstfdm}), which, with particle masses of $10^{-22}$ eV, can produce solitonic cores $\sim 1$ kpc in size, and is able to reproduce the cores observed in the dwarfs of the Local Group \citep{Hui_2017}. FDM is comprised of ultralight bosons that form a Bose-Einstein Condensate and is described by a complex scalar field. In this model, structure formation is inhibited below the de Broglie wavelength but behaves similarly to CDM on larger scales \citep{firstfdm}. Some models of ultra-light DM include interactions, which vary from single-field with SIs (\citealt{peebles2000} and \citealt{Rindler_Daller_2012}) to multi-field with non-trivial couplings (\citealt{matos2000}, \citealt{Bettoni_2014}, and \citealt{berezhiani15}). In this paper, we explore scalar field dark matter (SFDM; \citealt{lee1996}) as a modified model of $\Lambda$CDM. By modifying the microphysics of the dark matter, this model, just like FDM, diverges the most from CDM on small scales but behaves similarly to it on large scales, preserving the successful framework of the standard model \citep{sfdm2014}. However, in this case, the interactions between the particles give rise to a fluid pressure that produces halo cores. Comparing the rotation curves of the haloes formed in this cosmology with the ones found in dwarf galaxies in the local Universe, SFDM-haloes are able to simultaneously address the TBTF and core-cusp problems when a strong enough repulsive SI is also present \citep{sfdm2021}. A more detailed explanation of this model is addressed later in this paper. In order to solve the small-scale problems, \cite{sfdm2021} require a characteristic length scale for the model of $\gtrsim 1$ kpc. \cite{sfdm2014, sfdm2017} constrain this value to $\lesssim 5$ kpc based on observations from the CMB temperature anisotropy power spectrum. These limits correspond to self-coupling strengths in the range $1.6\times 10^{-18} \lesssim g/m^2c^4 \lesssim 4\times 10^{-17}$eV$^{-1}$cm$^3$ and require a particle mass larger than $\sim (2 - 10)\times 10^{-22}$ eV/$c^2$. \cite{shapiro}, however, find, by using constraints on FDM as a proxy for SFDM, that the characteristic length scale of SFDM should be as low as 10 pc. \cite{hartman2022} find this value to be $<1$ kpc. These are, to our knowledge, the only constraints placed on the SFDM model described by \cite{sfdm2014}.
\par 
To understand and solve small-scale challenges to the $\Lambda$CDM paradigm, we need to study the smallest structures in the Universe. The most encouraging class of objects for this purpose is possibly ultra-faint dwarfs (UFDs). These galaxies are defined with an absolute magnitude of $M_V\gtrsim -8$ and have total luminosities below those of individual bright red supergiant stars. Their sizes are intermediate between typical globular clusters and low-luminosity dwarf spheroidal galaxies \citep{willman}. Ultra-faint dwarfs represent the extreme limit of the galaxy formation process: they have the lowest metallicities, oldest ages, smallest sizes, and smallest stellar masses of all known galaxies \citep{simon}. Unlike most of the larger systems, they have survived to the present day as relics of the early universe. These objects therefore present us with a unique window into the conditions prevalent at the time when the first galaxies were forming (see e.g. \citealt{Bovill_2011}, \citealt{Wheeler_2015}). Furthermore, they reside in the smallest dark matter haloes yet found and have almost negligible baryonic masses, which means that their dynamical mass–to–light ratios are the highest ever measured \citep{simon}. Additionally, baryonic feedback should not significantly influence the dark matter density profiles of UFDs \citep{orkney2021} since they do not have enough stars to have significant supernovae feedback \citep{penarrubia}. Because of all these arguments, UFDs have attracted a lot of attention in the last few years. The known number of these objects has been increasing due to photometric catalogues from imaging surveys, and most of them are Milky Way satellites. The density profiles and measured central densities of UFDs can thus help us to constrain the nature of dark matter and its properties, making these objects unprecedented laboratories for understanding the behaviour of dark matter on small scales (see e.g. \citealt{calabrese}, \citealt{Errani_2018}, \citealt{bozek2018}). \par
However, the faintness of UFDs makes it challenging to obtain large numbers of stellar line-of-sight velocities and thus high-precision density profiles. Antlia B (Ant B), a slightly brighter dwarf galaxy (M$_V = -9.7$, M$_\star \sim 8\times10^5$M$_\odot$), promises more data, while at this stellar mass it is still expected that any star-formation-induced core would be very small (e.g. \citealt{bullock}, \citealt{orkney2021}). This makes Ant B well suited for testing models of dark matter. We present here an in-depth analysis of the kinematics of Ant B \citep{bas2022} from MUSE-Faint \citep{bas2020}, a survey of UFDs with the Multi Unit Spectroscopic Explorer (MUSE) \citep{bacon10}. With these data, which cover a wide range of radii, and adopting the Jeans analysis code \textsc{GravSphere} \citep{read2017}, we are able to derive the dark matter density profile of Ant B and, ultimately, to place constraints on the properties of SFDM. \par 
In Sect. \ref{antlia} we describe the observations (Sect.~\ref{sec:obs}), summarise the data reduction (Sect.~\ref{sec:reduction}), and describe the selection of the members of our dwarf galaxy, showing the determination of line-of-sight velocities from the spectra as well as its velocity distribution (Sect.~\ref{sec:selection}). This is followed in Sect.~\ref{methods} by the models of dark matter used in this paper (Sect.~\ref{sub:models}), by a brief description of \textsc{GravSphere} (Sect.~\ref{sec:gravsphere}), the analysis tool used, and by the priors used for each model (Sect.~\ref{sub:priors}). We continue in Sect.~\ref{results} with our results for the constraints on the properties of the studied dark matter models (Sect.~\ref{sec:constraints}) and for the recovered dark matter density profiles (Sect.~\ref{sec:density}). We end with a discussion in Sect.~\ref{discussion} and our conclusions in Sect.~\ref{conclusions}. In Appendix~\ref{app:parametrization} we show the numerical implementation of the SFDM model, presented in Sect.~\ref{sub:models}. The fitted parameters by \textsc{GravSphere} are presented in Appendix~\ref{app:gravspherefits}, followed by the relevant plots to discuss the robustness of the dark matter constraints obtained in Appendix~\ref{app:robustnessplots}.

 
\section{Antlia B} \label{antlia}
Antlia B was discovered by~\citet{sand2015} as a companion to NGC 3109 using DECam at Cerro Tololo International Observatory. Deep optical \textit{Hubble Space Telescope} (\textit{HST}) images that followed the ground-based discovery of this dwarf have been presented by \citet{hargis2020}. The photometric data obtained in both of these studies provided the measurements of the structural properties that are adopted in the present paper; these are listed in Table~\ref{tab:properties}. \par

\begin{table}[H]
\caption{Properties of Antlia B.}
\label{tab:properties}
\begin{center}
\begin{tabular}{lcc}
\hline
{\textbf{Parameter}} & {\textbf{Value}}        &  {\textbf{Source}} \\ \hline 
RA (deg)                       & $147.2337$ & \citet{sand2015}    \\ 
Dec. (deg)                  & -$25.9900$ & \citet{sand2015}    \\ 
$D$ (Mpc)                              & $1.35 \pm 0.06$          & \citet{hargis2020}  \\ 
$M_V$ (mag)                            & $-9.7\pm 0.6$            & \citet{sand2015}    \\ 
$r_h$ (pc)                        & $273\pm 29$              & \citet{sand2015}    \\ \hline
\end{tabular}
\end{center}

\footnotesize{\textbf{Notes.} (RA, Dec) are the coordinates of the centre of the galaxy; $D$ is the distance; $M_V$ is the absolute magnitude in the $V$-band; and $r_h$ is the half-light radius.}
\end{table}

Both studies agree that this dwarf has two distinct stellar populations: an old and metal-poor red giant branch population (>10 Gyr, [Fe/H]$\sim -2$) and a younger, more metal-rich population ($\approx 200-400$ Myr, [Fe/H]$\approx -1$). The theoretical isochrones for both of these populations are shown by \citet{sand2015} and \citet{hargis2020}, and there is no evidence of recent star formation ($\lesssim 10$ Myr). The star formation history measured by \citet{hargis2020} is compatible with the star formation seen in dwarf irregular galaxies in the Local Group: there is a constant growth in mass for the first $\approx 10$ Gyr and a rise in star formation in the last $\approx 2-3$ Gyr. \par
\citet{sand2015} obtain H$\alpha$ imaging of Ant B and show that the non-detection implies the lack of a stellar population <100 Myr in age. \par
These studies focus on photometry and do not have spectroscopic data, thus the dynamical properties of the stellar population have not been characterised. This is rectified by the present work. \par

\subsection{Observations}\label{sec:obs}
Ant B was observed with MUSE \citep{bacon10} on Unit Telescope 4 of the Very Large Telescope (VLT). MUSE is an optical wide-field spectrograph that uses the image slicing technique to cover a field of view of $1'\times 1'$ in wide-field mode with a spatial sampling of $0.2''\times0.2''$ \citep{bacon17}. The moderate spectral resolution (FWMH = $[2.4,3]$ $\AA$), broad wavelength range ($[4650,9300]$ $\AA$), good stability, and relatively large field of view permit the efficient acquisition of the spectra of very faint stars, making MUSE an ideal instrument to study compact UFDs \citep{bas2020}. \par
The 18 exposures total 4.5h, each one with an exposure time of 15 minutes, and were taken during the Guaranteed Time Observing (GTO, ESO ID 100.D-0807) runs between 11 and 15 February 2018 (9 exposures), between 14 and 18 March 2018 (6 exposures), and between 11 and 18 April 2018 (3 exposures). The observations followed a standard pattern of small dithers plus 90-degree rotation between each exposure. \par

\subsection{Data reduction}\label{sec:reduction}
The data reduction procedure used in this work follows the one described in \citet{bas2021}. In brief, we adopted the standard method for reducing MUSE data with the MUSE Data Reduction Software \citep{weilbacher2}, complemented with a bad-pixel table from \citet{bacon17}. Next, the produced data cube was post-processed with the Zurich Atmosphere Purge (version 2.0; \citealt{soto}) to remove residual sky signatures. \par
We used the public images of Ant B from the \textit{HST} (HST-GO-14078; PI: J. Hargis) in the V-band (F606W filter) and another in the I-band (F814W filter) in order to create a photometric catalogue using \textsc{SExtractor} \citep{sextractor}. 

\subsubsection*{Spectra}
We extracted spectra of the stellar sources in the MUSE cube using the \textsc{PampelMuse} code \citep{kamman}. The positions and magnitudes of the stellar sources, used as input to PampelMuse, were extracted from the \textit{HST} catalogue previously created with \textsc{SExtractor}. In summary, \textsc{PampelMuse} determines an initial point spread function (PSF) in the MUSE data, which is modelled with an analytic Moffat profile. After identifying in the reference catalogue the sources for which it is feasible to extract spectra, the spectra are extracted by simultaneously fitting a PSF to all sources that have been identified as resolvable. Since our field of view is very crowded, it was necessary to change the default value of \texttt{apernois}, the parameter that defines the maximum fraction of contaminating flux by nearby sources inside a PSF aperture. This parameter was changed from 0.1 to 0.4, meaning that the maximum allowed contamination around a PSF source from a neighbouring source has to be smaller than this value. In total, 2514 spectra were extracted with \textsc{PampelMuse}.

\subsubsection*{Line-of-sight velocities}
The line-of-sight velocities were determined from the spectra collected from \textsc{PampelMuse} using \textsc{spexxy} \citep{husser}. This software allows the determination of stellar properties by comparing spectra with synthetic spectra. \par
We adopted the same grid of synthetic \textsc{Phoenix} spectra (the Göttingen Spectral Library; \citealt{husser-phoenix}) used by \citet{husser16} for the globular cluster NGC 6397, by \citet{roth18} for the nearby galaxy NGC 300, and \citet{bas2020} for the UFD Eridanus II. The models are calculated on a grid of effective temperature, $T_{\text{eff}}$, logarithm of surface gravity, $\log g$, iron abundance, [Fe/H], and alpha-element abundance, [$\alpha$/Fe], and for each model the line broadening, $\sigma$, and line-of-sight velocity, $v_{\text{LOS}}$, can be fit. We fixed [$\alpha$/Fe] to zero (solar) since the quality of the spectra is not high enough to differentiate between different values of this parameter. The free parameters are determined by a weighted non-linear least squares minimisation against high-resolution synthetic spectra. 
The uncertainties determined by \textsc{spexxy} underestimate the scatter in the velocities below a signal-to-noise ratio (SNR) of 5 \citep{kamann2016} so we only retain spectra with a SNR above or equal to this for the subsequent analysis. Out of the 2514 spectra extracted with \textsc{PampelMuse}, 141 satisfied this criterion. \par 
\textsc{spexxy} was able to determine a velocity and metallicity, as well as their associated uncertainties, for 131 of the 141 spectra with SNR > 5. For the other 10 spectra, the fit failed to converge, and they were left out.
\par To obtain photometry for the stars, we cross-matched our catalogue with the  \cite{hargis2020} catalogue. All of our sources had counterparts in this catalogue.

\subsection{Selecting member stars}\label{sec:selection}
We discuss the selection of member stars in greater detail in the companion paper Brinchmann et al. (in prep.) but we summarise it here briefly. \par 
To clean up our list of possible members, we compared the \textit{HST} photometry against \textsc{parsec} \citep{parsec} isochrones. We adopted the metallicities and ages estimated by \citet{hargis2020} and found that all 131 member stars were consistent with at least one isochrone at the distance of Ant B. We also inspected all spectra manually to verify that they were all stellar spectra. \par
\begin{figure}[h]
  \centering
  \includegraphics[scale=0.5]{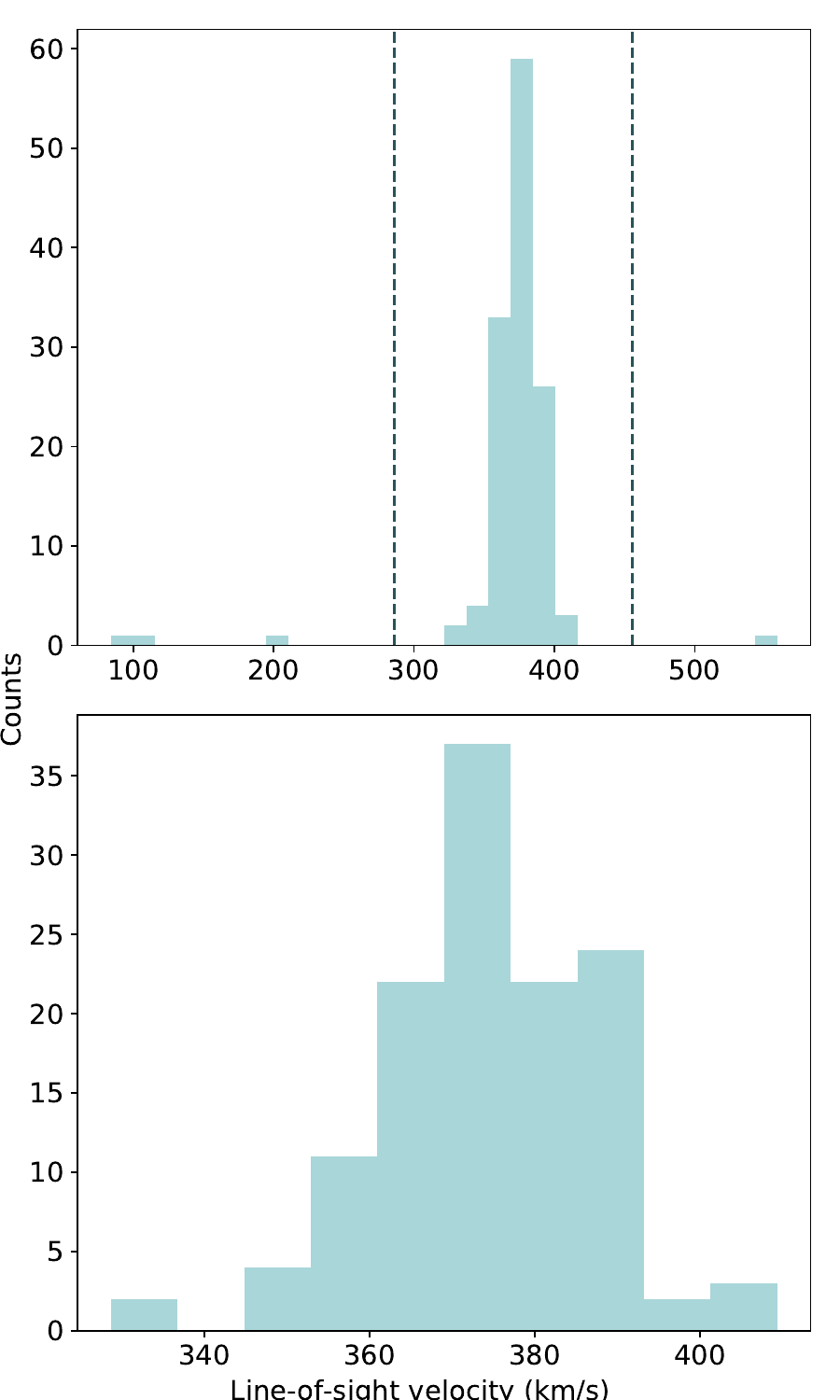}
    \caption{Histogram of the line-of-sight velocity for Ant B candidates. \textbf{Top}: All the candidates for which we have spectra (131 stars), with the velocity cut, $v = \mu \pm 2\sigma$, represented by the dashed lines. \textbf{Bottom}: After the velocity cut (127 stars).}%
    \label{fig:velcut}
\end{figure}

For the analysis here we also needed to exclude strong outliers in velocity. To do this, we calculated the mean ($\mu\approx 370.5$km s$^{-1}$) and standard deviation of the velocities of the 131 candidate stars ($\sigma \approx 42.4$ km s$^{-1}$) and excluded stars outside $\mu \pm 2\sigma$ (see Figure~\ref{fig:velcut}). This excluded four clear outliers in velocity, leading to a final sample of 127 member stars.

\begin{figure}[h]
   \centering
   \includegraphics[width=\columnwidth]{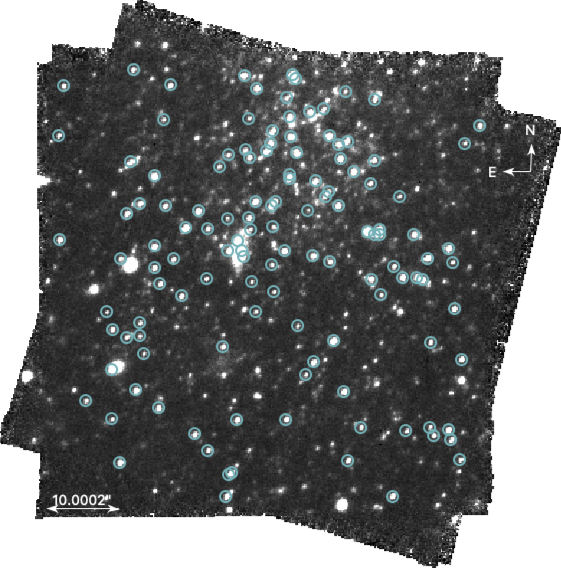}
   \caption{Image of the final cube (all the exposures described before were combined in the data reduction) with the 127 stars identified as Ant B members represented by the blue circles.}\label{fig:finalselection}
\end{figure}

As discovered by \cite{sand2015}, Ant B has \hi\ gas at a velocity of $v_{\text{H}_\text{I}} = 376\pm 2$ km s$^{-1}$. The good agreement in line-of-sight velocity is encouraging and confirms that the stars and \hi\ are associated. We then continued our analysis with 127 member stars of Ant B. These members are represented in Figure~\ref{fig:finalselection}. The velocities of these members have previously been presented by \citet{bas2022} and the resulting catalogue can be found there. 

\subsubsection*{Velocity distribution}
In order to determine the intrinsic mean value and dispersion of the velocity of Ant B, it is necessary to take into account the associated measurement uncertainties. To this purpose, we adopted a Markov Chain Monte Carlo (MCMC) approach (see e.g. \citealt{hargreaves94}, \citealt{martin15}) described in detail in \citet{bas2020}, which assumes that the velocity distribution of the stars can be modelled with a Gaussian. 
The global likelihood of the system is then
\begin{equation}
    \mathcal{L}(\mu,\sigma|v_i, \sigma_i) = \prod_i\left[\frac{1}{\sqrt{2\pi}\sigma_{\text{obs},i}}\exp\left(-\frac{1}{2}\left(\frac{v_i-\mu}{\sigma_{\text{obs},i}}\right)^2\right)\right],
    \label{eq:mcmcglobal}
\end{equation}
where $v_i$ is the velocity of star $i$, $\mu$ is the mean line-of-sight velocity of the system, and $\sigma_{\text{obs},i}^2 = \sigma^2 + \sigma_{i}^2$ is the observed velocity dispersion for star $i$.
We adopted a flat prior on $\mu$ between 285.7 and 455.3 km s$^{-1}$ to be consistent with our member selection. We also adopted a flat prior on $\sigma$ between 0 and 40 km s$^{-1}$. The results are robust to the choice of prior. \par
\begin{figure}[h]
   \centering 
   \includegraphics[width=\columnwidth]{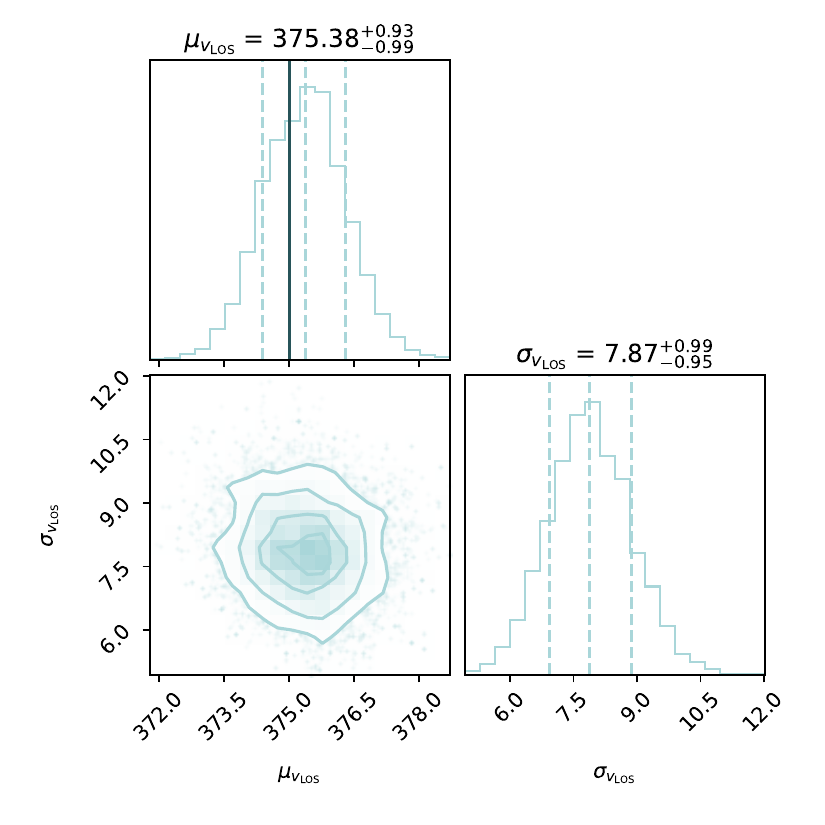}
   \caption[Corner plot for the MCMC velocity fit]{Corner plot for the MCMC velocity fit, using the 127 member stars of Ant B. The histograms along the diagonal represent the posterior distribution for each parameter: the mean value $\mu_U$ in the top panel and the dispersion $\sigma_U$ in the bottom right. The vertical dashed lines indicate the median and 68$\%$ confidence interval. The bottom-left panel represents the 2D posterior distributions of both of these parameters, with the contours corresponding to $0.5\sigma$, $1\sigma$, $1.5\sigma$, and $2\sigma$ confidence levels, where $\sigma$ is the standard deviation of the 2D distribution. The solid dark blue line in the top panel indicates the mean velocity of the \hi\ gas reported by \citet{sand2015}.}\label{fig:veldist}
\end{figure}

We obtained a mean value of $v_{\text{LOS}}=375.39^{+0.98}_{-0.97}$ km s$^{-1}$ and an intrinsic velocity dispersion of $\sigma_{v_{\text{LOS}}}=7.87^{+1.02}_{-0.98}$ km s$^{-1}$. The resulting corner plot is represented in Figure~\ref{fig:veldist}.
We see that there is good agreement between the mean velocity of the stars and the mean velocity of the \hi\ gas reported by \citet{sand2015}.


\section{Methods}\label{methods}
We start by describing the dark matter models tested in this paper in Sect. \ref{sub:models}. This is followed in~\ref{sec:gravsphere} by the presentation of the analysis tool (\textsc{GravSphere}; \citealt{read2017}, \citealt{draco}, \citealt{genina}, \citealt{read2019}, \citealt{collins2021}) used to constrain the dark matter profiles and their microphysical properties.

\subsection{Models of dark matter density profiles}\label{sub:models}
\subsubsection*{Cold dark matter}

We adopted the Navarro-Frenk-White (NFW, \citealt{nfw}) profile as the parametric description of cold dark matter density profiles, 
\begin{equation}
\rho_{\text{CDM}}(r) = \frac{\rho_0}{(r/r_s)(1+r/r_s)^2},
\label{eq:CDM}
\end{equation}
where the characteristic density, $\rho_0$, is defined by ${\rho_0=\rho_{\text{crit}}\Delta^3c_{200}g_c/3}$, and the scale radius, $r_s$, is defined by $r_s = r_{200}/c_{200}$. The critical density, $\rho_{\text{crit}}$, is defined by $\rho_{\text{crit}} = 3H^2/8\pi G$, where $H$ is the Hubble constant and $G$ the gravitational constant. Finally, the virial radius, $r_{200}$, is defined by
\begin{equation}
r_{200} = \left[\frac{3}{4}M_{200}\frac{1}{\pi200\rho_{\text{crit}}}\right]^{1/3},
\label{eq:r200}
\end{equation}
where $M_{200}$ is the virial mass enclosed in the virial radius. ${\Delta\sim200}$ is an overdensity constant relative to the background matter density, $c_{200}$ a concentration parameter, and $g_c$ is defined as
\begin{equation}
g_c = \displaystyle\frac{1}{\log(1+c_{200})-\frac{c_{200}}{1+c_{200}}}.
\label{eq:gc}
\end{equation}
\subsubsection*{Scalar field dark matter}

Scalar field dark matter (\citealt{sfdm2014}) is comprised of ultralight bosons and is described by the coupled Schrödinger-Poisson equations. This model is an alternative to $\Lambda$CDM and in this case it leads to a Bose-Einstein condensate and quantum superfluid. Just like FDM, this model's behaviour approaches the CDM model on large scales and differs the most on small scales. The main difference between SFDM and FDM is that SFDM also includes a repulsive SI \citep{sfdm2021}. \par
While CDM is described by the collisionless Boltzmann equation, SFDM obeys the non-linear Schrödinger equation, rewritten in terms of quantum hydrodynamics equations for the conservation of mass and momentum. \par
In the SFDM model, there are two length scales that characterise the scale below which the structure is suppressed. On the one hand we have the de Broglie wavelength $\lambda_{\text{deB}} = h/mv$, which is characteristic of the FDM model (in the case in which SI is not present - the free field limit) and which depends on the boson mass, $m$, and its characteristic velocity, $v$. On the other hand, we have $\lambda_{\text{SI}}$, resulting from the presence of a repulsive SI. \par 
In FDM, structure formation is suppressed on scales below $\lambda_{\text{deB}}$. In the presence of the repulsive SI with self-coupling strength $g$, $\lambda_{\text{SI}} \propto \sqrt{g/m^2}$, and if the interaction is strong enough that $\lambda_{\text{SI}} \gg \lambda_{\text{deB}}$, structure is suppressed below $\lambda_{\text{SI}}$.
In this regime, the Thomas-Fermi (TF) regime, $\lambda_{\text{deB}}$ is much smaller than the characteristic length scale of the repulsive SI, $R_{\text{TF}}$, given by
\begin{equation}
R_{\text{TF}} = \pi\sqrt{\frac{g}{4\pi G m^2}}.
\label{eq:rtf}
\end{equation}
This quantity is a physical constant of SFDM since it is fixed by the particle parameters ($m, g$) in their combination $g/m^2$, called the SI strength parameter. $R_{\text{TF}}$ is the relevant scale below which the Thomas-Fermi regime of SFDM differs from CDM. To avoid the small-scale structure problems of the standard model discussed before, either $\lambda_{\text{deB}}$ for the FDM model, or $\lambda_{\text{SI}}$ for the SFDM-TF model, should be $\sim 1$ kpc.\par
The density profile of the haloes formed in SFDM-TF was fitted by \citet{sfdm2021} and is given by 
\begin{equation}
\rho_{\text{SFDM}} =
\begin{cases}
     \rho_c\sinc(\pi r/R_{\text{TF}}) & r \leq \alpha R_{\text{TF}} \\
     \rho_{\text{CDM}} & r \geq \alpha R_{\text{TF}}
    \end{cases},
\label{eq:sfdm}
\end{equation}
where $\sinc(x)=\sin(x)/x$, $\rho_c$ is the central density, and $\alpha$ is a given fraction of $R_{\text{TF}}$ determined by requiring mass conservation and density continuity at $r = \alpha R_{\text{TF}}$. As is clear from this equation, SFDM-TF behaves just like the CDM model on scales larger than $R_{\text{TF}}$.

\subsection{\textsc{GravSphere}} \label{sec:gravsphere}
To measure the dark matter density profiles of Ant B, we used the updated version of \textsc{GravSphere} Jeans modelling code\footnote{The newest version of this code is available for download at \url{https://github.com/justinread/gravsphere} \citep{collins2021}.}. A detailed explanation of its implementation can be found in \citet{read2017} and \citet{draco}, which we briefly mention here for the reader's convenience. \par 
\textsc{GravSphere} solves the Jeans equation \citep{jeans} for our member stars while assuming that the stellar system is spherical, non-rotating, and in a steady state, given by 
\begin{equation}
\frac{1}{\nu_\star}\pdv{}{r} \left(\nu_\star\sigma_r^2\right)+\frac{2\beta(r)\sigma_r^2}{r} = -\frac{GM(<r)}{r^2},
\label{eq:jeans}
\end{equation}
where 
\begin{equation}
\sigma_r^2 = \langle v_r^2\rangle - \langle v_r\rangle^2 \text{ with }\langle v_r^n\rangle = \int v_r^n f\dd^3\textbf{v},
\label{eq:sigmar2}
\end{equation}
and $M(<r)$ is the total cumulative mass as a function of the radius, $r$. The tracer number density $\nu_\star$ characterises the radial density profile of a population of massless tracers (in our case, stars moving in a galaxy) that move in the gravitational potential of its mass distribution $M(r)$, modelled with three \citet{plummer} profiles
\begin{equation}
\nu_\star(r) = \sum_{j=1}^3 \frac{3M_j}{4\pi a^3_j}\left(1+\frac{r^2}{a^2_j}\right)^{5/3}
\label{eq:tracerdensity},
\end{equation}
with masses $M_j$ and scale length $a_j$ for each individual component. This enables the recovering of the density profile, $\rho(r)$, and the velocity anisotropy profile, $\beta(r)$, of the studied stellar systems \citep{read2017}. The velocity anisotropy, $\beta(r)$, is defined as
\begin{equation}
\beta(r)= \beta_0+\frac{\beta_\infty-\beta_0}{1+(r_0/r)^\eta},
\label{eq:ani}
\end{equation}
where $\beta_0$ is the central value of the anisotropy, $\beta_\infty$ is the
value at infinity, $r_0$ is a transition radius, and $\eta$ is the steepness of the transition. To avoid infinities, a symmetrised version of $\beta(r)$ is used,
\begin{equation}
\Tilde{\beta}(r)= \frac{\sigma_r(r)-\sigma_t(r)}{\sigma_r(r)+\sigma_t(r)}=\frac{\beta(r)}{2-\beta(r)}.
\label{eq:anitil}
\end{equation}
$\tilde{\beta} = 0$ corresponds to an isotropic velocity dispersion, $\tilde{\beta} = 1$ to a fully radial dispersion, and $\tilde{\beta} = -1$ to a fully tangential distribution. \par
\textsc{GravSphere} relies on higher order moments of the velocity distribution via the fourth order Virial Shape Parameters (VSPs) \citep{vsps} to partially break the degeneracy that exists between the radial velocity dispersion (and therefore the cumulative mass distribution) and the velocity anisotropy. To bin the data in bins of the projected radius, $R$, we used an algorithm called \textsc{Binulator} \citep{collins2021}. \par
After performing the \textsc{Binulator} routine, we applied our adaptation of \textsc{GravSphere}\footnote{The alternative dark matter models implemented are available for download at \url{https://github.com/marianajulio/alternative_models_for_gravsphere}.} to the Ant B data in order to get our dark matter density profile and, consequently, the dark matter constraints we were looking for. \textsc{GravSphere} solved the Jeans equation for the projected velocity dispersion. It also fitted the two VSPs, with the initial guesses being the profiles estimated by \textsc{Binulator}. \par
\textsc{GravSphere} uses the ensemble sampler \textsc{Emcee} \citep{emcee} to fit the model to the data. Each individual Markov chain (walker) communicates with the other walkers at each step, contrary to the classic Metropolis-Hastings algorithm, and hence it allows the chains to sample the posterior distribution more efficiently \citep{genina}. The number of walkers and steps used depends on the model considered since some models require more time to achieve convergence. The first half of the steps generated are always discarded as a conservative burn-in criterion.

\subsection{Priors}\label{sub:priors}
The standard priors that \textsc{GravSphere} requires were kept to their default values, which we summarise here for convenience. We used priors on the symmetrised velocity anisotropy of $ -1 < \Tilde{\beta}_0 < 1$, $-1 < \Tilde{\beta}_\infty < 1$, $-1 < \log_{10}(r_0/\mathrm{kpc}) < 0$, and $1 < \eta < 3$. We used a flat prior on the stellar mass of $6 < M_\star/(10^{5}\mathrm{M}_\odot) < 10$. \par
For the CDM model, the only quantities that we need to know in order to calculate all the other parameters are $M_{200}$ and $c_{200}$, since $r_s = r_{200}/c_{200}$ and $r_{200}$ can be estimated by Equation~\ref{eq:r200} (which depends only on $M_{200}$ and some constants), and $g_c$ by Equation~\ref{eq:gc} (which only depends on $c_{200}$). \par
Although we do not know the exact values of these parameters for the systems that we aim to study, the smallest dwarf galaxies show $M_{200}\approx 10^8$M$_\odot$ \citep{bullock}. Such a limit is also theoretically expected, since below this mass, galaxy formation becomes extremely inefficient due to hindered atomic cooling (e.g. \citealt{finkelstein_conditions_2019}). To be considered small-scale, the maximum value that $M_{200}$ can take is approximately $10^{11}$M$_\odot$ \citep{bullock}. For this reason, we used a generous flat prior on $M_{200}$ ranging from $M_{200}= 10^{7.5}$M$_\odot$ to $M_{200}=10^{11.5}$M$_\odot$ that encompasses the mass-range of all dwarf galaxies \citep{Collins_2022}. \par  
The concentration parameter $c_{200}$ is related to $M_{200}$. We used the mass-concentration relation for cold dark matter derived by \citet{c200m200}, based on observations of strong lenses, to determine a prior for the concentration parameter that was proper for a galaxy with the mass of Ant B. Since the covered mass range of this relation ($M_{200} = [10^6,10^{10}]$M$_\odot$) is very similar to what we use, it seemed appropriate to use the resulting values of the concentration parameter. They constrained the concentration at $z=0$ with a $95\%$ confidence of $c = 15^{+18}_{-11}$ for $M_{200} = 10^7$M$_\odot$ and $c = 10^{+14}_{-7}$ for $M_{200} = 10^9$M$_\odot$. Since we were exploring not only CDM but also SFDM, we assumed that a generous flat prior for the concentration parameter was then $c = [1, 50]$. These values are described in the two top rows of Table~\ref{tab:priors}. \par

Since this model has a straightforward density and mass profile, it does not need a high number of walkers and steps to converge. It was found that using $n_{\text{walkers}} = 250$ and $n_{\text{steps}} = 25000$ was sufficient to achieve convergence, making it the fastest model to run. Several runs were made with different values for these parameters, and this choice did not affect the final results.\par 
The parameters of the SFDM density profile that require priors are $M_{200}$, $c_{200}$, and $R_{\text{TF}}$. For $M_{200}$ and $c_{200}$, we adopted the same priors as for the CDM profile, but we note that it has not been established that the mass-concentration relation established for CDM applies to SFDM. For that reason, we kept the priors quite broad. $R_{\text{TF}}$ was chosen to take into account the possible values of this parameter discussed by~\citet{sfdm2021}. This parameter is zero when a core is not present and the density profile follows the CDM model and needs to be $\gtrsim 1$ kpc in order to solve the core-cusp problem and the TBTF problem for Local Group dwarfs. However, $R_{\text{TF}} \lesssim 5$ kpc is required based on observations from the CMB temperature anisotropy power spectrum if we consider dark matter as SFDM \citep{sfdm2021}, so our flat prior on $R_{\text{TF}}$ varied from 0.001 kpc to 5 kpc.

\begin{table}[H]
\caption{Priors used for the discussed dark matter models. CDM only uses the top two rows, while SFDM uses all the priors.}
\label{tab:priors}
\begin{center}
\begin{tabular}{lcc}
\hline
{\textbf{Parameter}} & {\textbf{Min.}} & {\textbf{Max.}}          \\ \hline
$\log_{10}(M_{200}$/M$_\odot)$     & 7.5 & 11.5     \\ 
$c_{200}$                          & 1 & 50     \\ 
$R_{\text{TF}}$ (kpc)              & 0.001 & 5  \\ \hline
\end{tabular}
\end{center}
\end{table}

Contrary to the CDM model, the SFDM model has to obey certain conditions to arrive at a physical solution for the density and mass profile. The implemented numerical parametrization of this model is discussed in detail in Appendix~\ref{app:parametrization}. Although $\alpha$ and $\rho_c$ are easily computed, all $r$ must be evaluated in order to choose which calculation to perform next: either the density and mass profile associated with the TF regime, or the ones associated with the CDM model. Furthermore, to calculate the mass profile associated with the TF regime, an integral must be solved. Moreover, this model needs a higher number of walkers and steps to converge than did the CDM model; we used $n_{\text{walkers}} = 500$ and $n_{\text{steps}} = 10^5$ to get robust results.


\section{Results}\label{results}
Using \textsc{GravSphere}, discussed in Sect.~\ref{sec:gravsphere}, we sampled the parameter spaces of our implemented dark-matter density models, examined in Sect.~\ref{sub:models}.
To do this, we used the kinematical measurements of Ant B, discussed in Section~\ref{antlia}. 
Here, in Sect.~\ref{sec:constraints}, we present the obtained dark matter constraints and, in Sect.~\ref{sec:density}, the recovered density profiles as well as their comparison. \par
\begin{figure}[H]
   \centering 
   \includegraphics[width=\columnwidth]{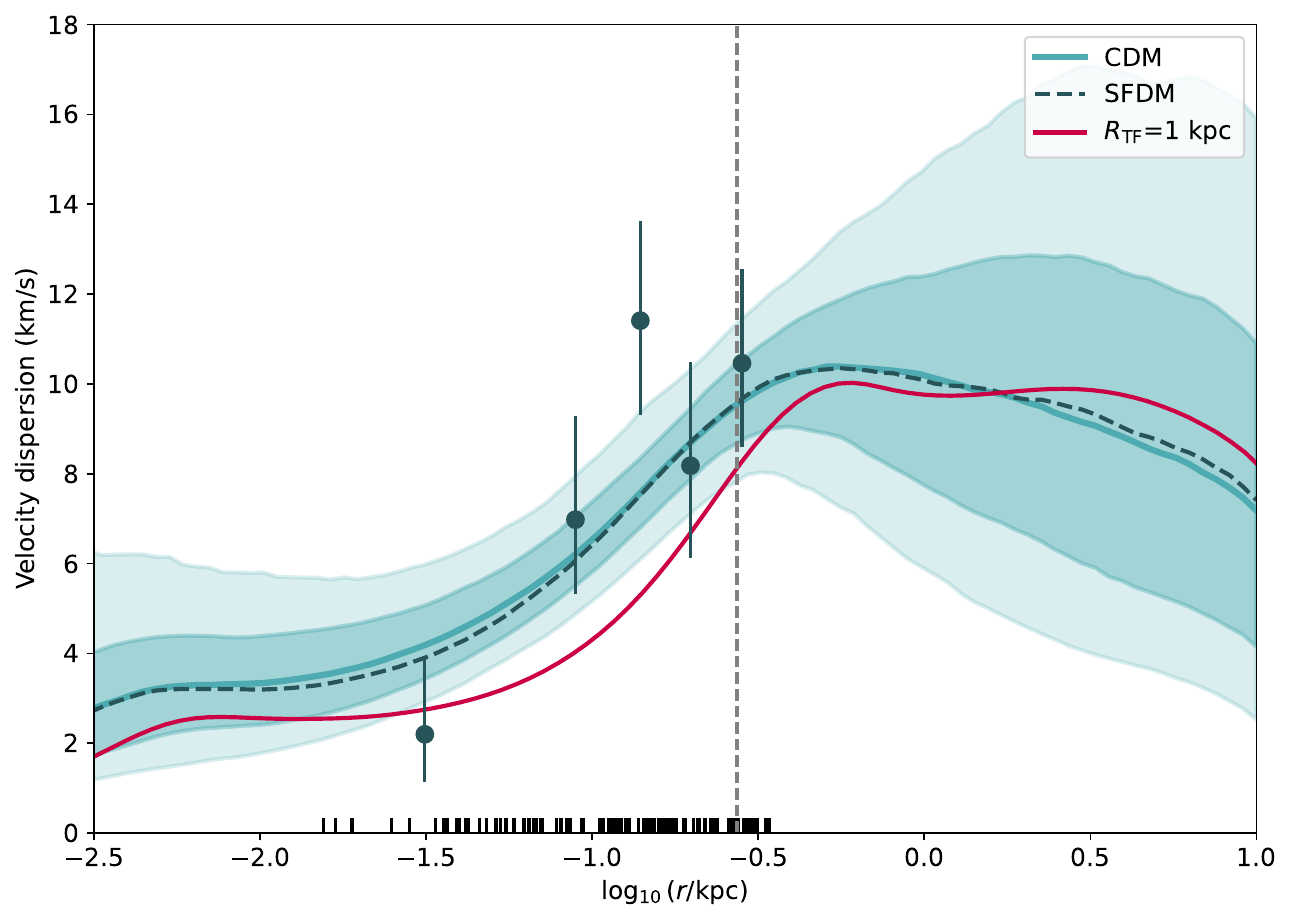} \\
   \caption[]{Velocity dispersion of Ant B. The dark blue points represent the binned velocity dispersion with the associated uncertainties. The best fit from \textsc{GravSphere} for the CDM model is shown as a solid light blue line, with the light blue shaded regions showing the 68$\%$ and 95$\%$ confidence intervals for the CDM model. The dashed dark blue line represents the best fit for the SFDM model. Since the best fit for both models is so similar, the confidence intervals for the SFDM model are omitted for clarity. The half-light radius is represented by the vertical dashed line and the bottom marks represent the projected radii of the members of Ant B. The solid red line represents the shape that the velocity dispersion would have if $R_\mathrm{TF} = 1$ kpc, keeping the other parameters with the same values as the best fit for the SFDM profile.}\label{fig:veldisp_recovered}
\end{figure}

 
To achieve this goal, we need the kinematic and photometric data of Ant B. \textsc{Binulator} starts by fitting the photometric data in order to get the light profile. In our case, we made a mock photometric catalogue by drawing 10,000 sample photometric positions from the exponential distribution from \cite{sand2015} and used them as photometric observations in \textsc{Binulator}. Next, the 127 member stars of Ant B were binned radially from the centre of Ant B (listed in Table~\ref{tab:properties}) into 5 bins, each of them having $\sim 25$ sources. The binned data that \textsc{GravSphere} uses is described in Table~\ref{tab:bin}. Figure~\ref{fig:veldisp_recovered} shows the projected radii for all the stars, their position after the binning routine, and the velocity dispersion associated with each bin in dark blue.

\begin{table}[H]
\caption[Kinematic data of Ant B after the binning routine performed by \textsc{Binulator}, as used by \textsc{GravSphere}.]{Kinematic data of Ant B after the binning routine performed by \textsc{Binulator}, as used by \textsc{GravSphere}. The bins are represented by the mean projected radius of the stars in each bin.}
\label{tab:bin}
\begin{center}
\begin{tabular}{cc}
\hline
{\textbf{Radius} (kpc)} & {\textbf{Velocity dispersion} (km s$^{-1}$)} \\ \hline
0.031      & $2.20^{+1.71}_{-1.06}$     \\ 
0.089      & $6.98^{+1.66}_{-2.30}$       \\ 
0.140      & $11.41^{+2.10}_{-2.23}$       \\ 
0.198       & $8.18^{+2.05}_{-2.32}$       \\ 
0.283     & $10.46^{+1.87}_{-2.09}$   \\ \hline
\end{tabular}
\end{center}
\end{table}

Our best-fitting CDM and SFDM models are shown in Figure~\ref{fig:veldisp_recovered}, compared with the stellar velocity dispersion data. Both models fit the data reasonably well but do not reproduce all features. In particular, both models have difficulties in fitting the low central velocity dispersion when they are anchored at the half-light radii. This figure also shows that, as expected, we lack constraining power where there is no data, both at small ($\sim$ 10 pc) and large radii ($\gtrsim$ 1 kpc). The recovered fits from \textsc{GravSphere} for the VSPs and the surface brightness profile can be seen in Figure~\ref{fig:vsps_surfb} of the Appendix.

\subsection{Dark matter constraints}\label{sec:constraints}
\begin{figure}[h]
   \centering
   \includegraphics[width=\columnwidth]{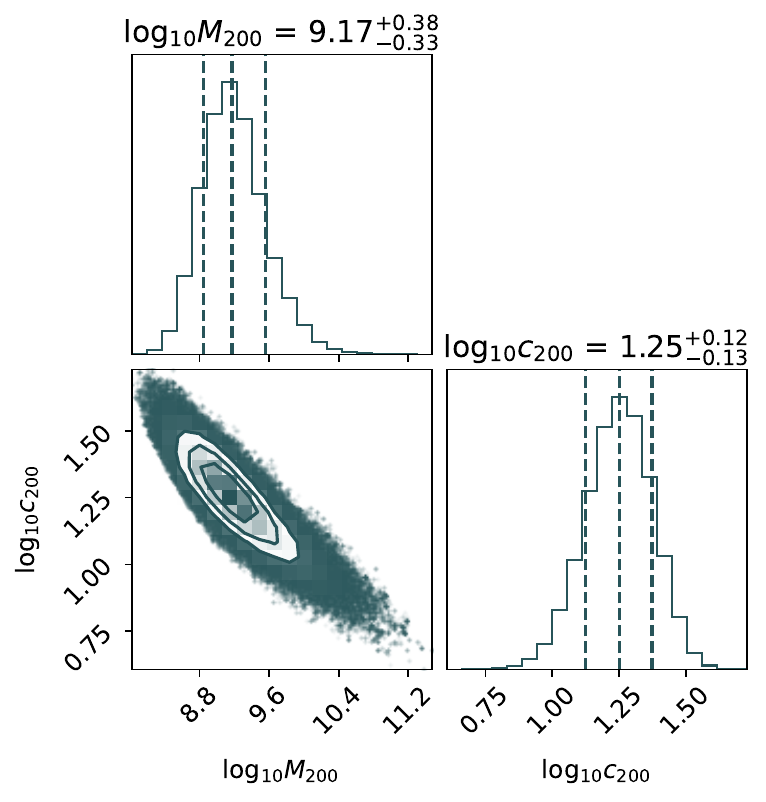}
   \caption[Constraints on the CDM profile for Ant B]{Constraints on the CDM profile for Ant B. The histograms along the diagonal represent the posterior distribution for each parameter: the virial mass, $M_{200}$, in M$_\odot$ and the concentration parameter $c_{200}$.
   Their units are omitted for clarity. The vertical dashed lines indicate the median and 68$\%$ confidence interval. The bottom left panel represents the 2D posterior distribution of these parameters, with the contours corresponding to the $0.5\sigma$, $1\sigma$, $1.5\sigma$, and $2\sigma$ confidence levels, where $\sigma$ is the standard deviation of the 2D distribution.}\label{fig:cornerCDM}
\end{figure}

\begin{figure}[h]
   \centering 
   \includegraphics[width=\columnwidth]{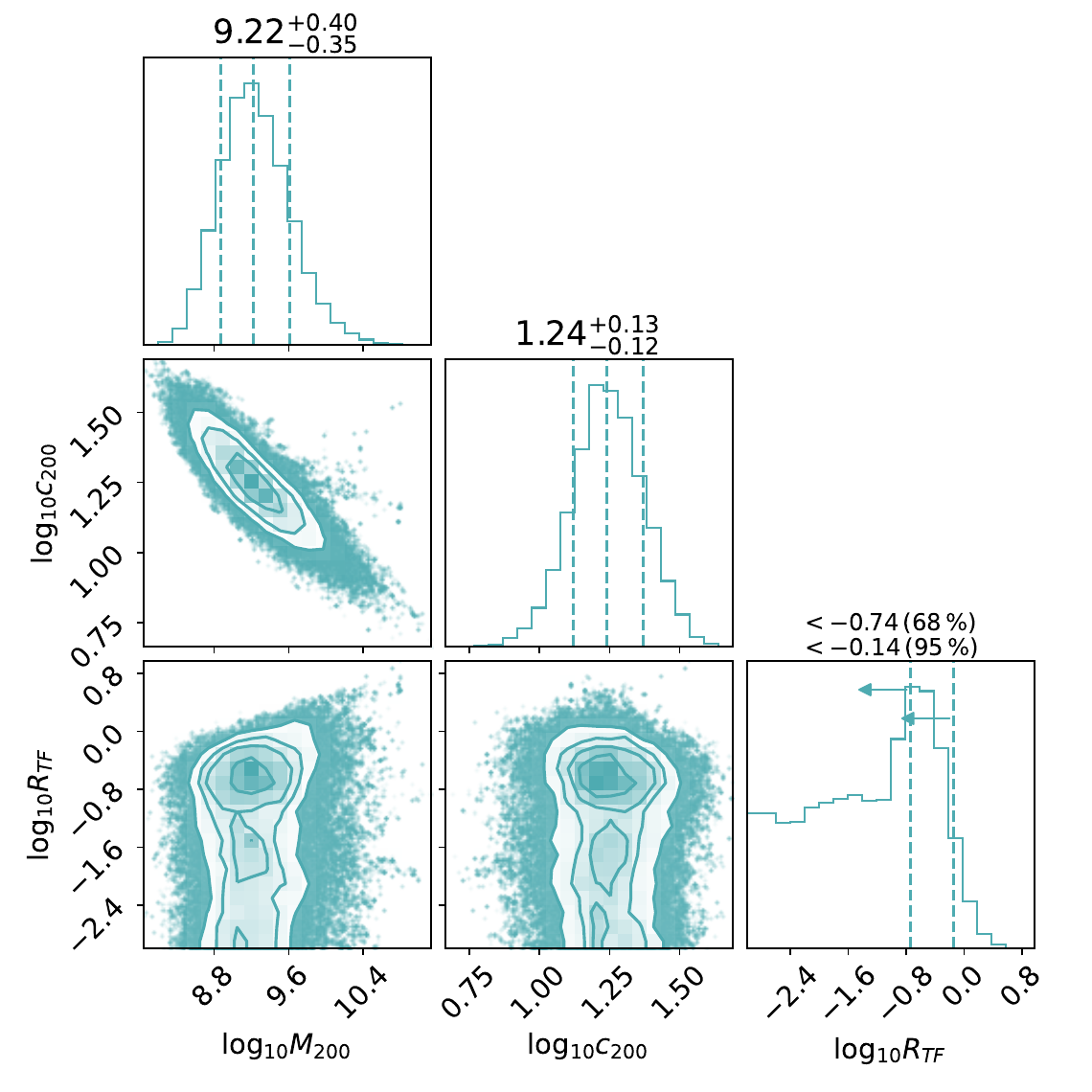}
   \caption[Constraints on the SFDM profile for Ant B.]{Constraints on the SFDM profile for Ant B. The histograms along the diagonal represent the posterior distribution for each parameter: the virial mass, $M_{200}$ in M$_\odot$, the concentration parameter $c_{200}$, and the characteristic length scale of the repulsive SI $R_{\text{TF}}$ in kpc. Their units are omitted for clarity. The vertical dashed lines indicate the median and 68$\%$ confidence interval (without arrows) and the $68\%$ and $95\%$ confidence limits (upper and lower arrows, respectively). The other panels represent the 2D posterior distributions of these parameters, with the contours corresponding to the $0.5\sigma$, $1\sigma$, $1.5\sigma$, and $2\sigma$ confidence levels, where $\sigma$ is the standard deviation of the 2D distribution.}\label{fig:cornersfdm}
\end{figure}

The constraints of the physical parameters of the CDM and SFDM models are displayed in Figures~\ref{fig:cornerCDM} and \ref{fig:cornersfdm}, respectively. The constraints in the computational parametrizations for the SFDM are shown in Figure~\ref{fig:corner_SFDM_rho} of the Appendix. To calculate the marginalised constraints, we took the median and quartiles of the chains after discarding samples with a $\chi^2$ larger than 10 times the minimum $\chi^2$. \par 

For the CDM profile we find a virial mass of ${M_{200}=10^{9.17^{+0.38}_{-0.33}}} \approx 1.48^{+2.29}_{-0.79}\times 10^9$ M$_\odot$ and a concentration parameter of ${c_{200}=10^{1.25^{+0.12}_{-0.13}}} \approx 17.78^{+5.66}_{-4.60}$. These values are summarised in Table~\ref{tab:results}. \par

For the SFDM profile, we find a virial mass of ${M_{200}=10^{9.22^{+0.40}_{-0.35}}} \approx 1.66^{+2.51}_{-0.92}\times 10^9$ M$_\odot$ and a concentration parameter of ${c_{200}=10^{1.24^{+0.13}_{-0.12}}} \approx 17.38^{+6.06}_{-4.20}$. Clearly, the SFDM and CDM profile fits are in good agreement for these parameters. We can only present the characteristic length scale of the repulsive SI as an upper limit. The posterior distribution that we get for $R_{\text{TF}}$ contains values ($R_{\text{TF}} \sim 0.63$ pc) that are smaller than the projected radii of our innermost tracer ($1.84$ pc), leading to a lack of constraining power at the lower end of the range of $R_{\text{TF}}$. We find $R_{\text{TF}}$ (kpc) $< 10^{-0.74} \lesssim 0.18$ at the $68\%$ confidence level and $R_{\text{TF}}$ (kpc) $< 10^{-0.14} \lesssim 0.72$ at the $95\%$ confidence level. These values are summarised in Table~\ref{tab:results}. \par 

\begin{table*}
  \centering
  \caption{Summary of the Bayesian evidence for CDM and SFDM, as well as the marginalised posterior estimates obtained for the free parameters of both models.}
  \begin{tabular}{l|cc|lrccrr}
  \hline
   \multicolumn{1}{l|}{\textbf{Model}} &
   \multicolumn{1}{c}{$\mathbf{\log_{10}}\textbf{(\textit{Z})}$} &
   \multicolumn{1}{c|}{$\mathbf{\Delta\log_{10}}$\textbf{(\textit{Z})}} &
    \multicolumn{1}{l}{\textbf{Parameter}} &
    \multicolumn{1}{c}{\textbf{Median}} &
    \multicolumn{1}{c}{\textbf{2.5\%}} &
    \multicolumn{1}{c}{\textbf{16\%}} &
    \multicolumn{1}{c}{\textbf{84\%}} &
    \multicolumn{1}{c}{\textbf{97.5\%}} \\
     \hline
  \multirow{2}{*}{\textbf{CDM}} & \multirow{2}{*}{$-124.83$} & \multirow{2}{*}{$0.00$} & $M_{200}/(10^9$ M$_\odot)$ & $1.48$ & $1.12$ & $0.79$ & $2.29$ & $8.75$ \\
  & & & $c_{200}$ & $17.78$ & $8.01$ & $4.60$ & $5.66$ & $12.42$ \\
\hline 
  \multirow{3}{*}{\textbf{SFDM}} & \multirow{3}{*}{$-125.49$} & \multirow{3}{*}{$-0.66$} & $M_{200}/(10^9$ M$_\odot)$ & $1.66$ & $1.28$ & $0.92$ & $2.51$ & $9.82$ \\ 
  & & & $c_{200}$ & $17.38$ & $7.60$ & $4.20$ & $6.06$ & $13.52$ \\ 
  & & & $R_{\text{TF}}$ (pc) & & & & $\lesssim 180$ & $\lesssim 720$ \\  \hline
  \end{tabular}
  \label{tab:results}
\end{table*}

\subsection{Comparison between profiles}\label{sec:density}
\begin{figure}[h]
   \centering
   \includegraphics[width=\columnwidth]{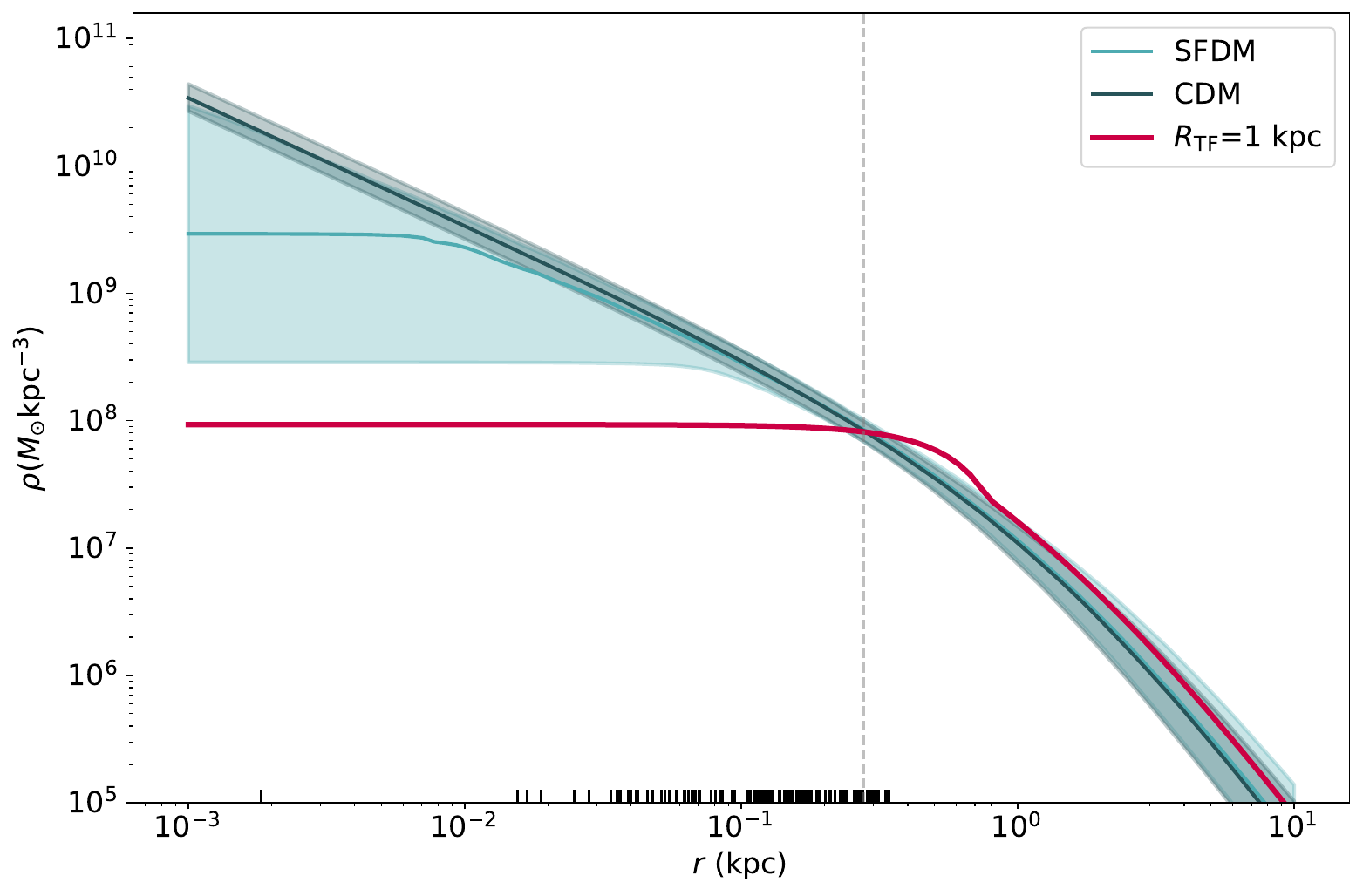}
   \caption[Recovered DM density profiles of Ant B.]{Recovered DM density profiles of Ant B for the adopted models. The solid lines represent the median density profile: the CDM model is represented in dark blue and the SFDM model in light blue. The filled areas represent the $68\%$ confidence interval. The solid red line represents the shape of the same density profile if $R_{\text{TF}} = 1$ kpc. The vertical dashed line indicates the half-light radius and  the bottom marks represent the projected radii of the members of Ant B.}\label{fig:dens}
\end{figure}

The density profiles were generated by drawing 1000 random samples from the remaining samples after the cut was performed on $\chi^2$. In Figure~\ref{fig:dens} we show the recovered density profiles for the three models adopted as a function of the radius. These density profiles are represented by the median density of the random samples and their $68\%$ confidence interval at every radius. \par
The SFDM model agrees with the CDM model within the uncertainties and their agreement is almost perfect at large radii. The uncertainties of the SFDM allow for the CDM model even at the smallest radii, when $R_{\text{TF}} \to 0$. Not surprisingly, the agreement of the discussed models is best at the radii at which the density of stars in our sample is highest. \par
The model that represents cold dark matter has the highest density and smallest uncertainties, while the SFDM model seems to prefer lower densities. However, the density estimates for SFDM are likely biased towards lower values because there is a physical limit set by CDM and the model only allows for that or lower values. \par
In Table~\ref{tab:results} we present the Bayesian evidence, $Z$, for both models and the decimal logarithm of the Bayes factor, $\Delta\log_{10}(Z)$, computed in relation to the model with the largest $Z$. This estimation follows the one described in \cite{bas2021, bas2022}, and we use \textsc{MCEvidence} \citep{mcevidence} for this purpose. We assume the prior probabilities of the models are equal and we take into account that these models have different degrees of freedom. We can compare the models by estimating the ratio of $Z$, with the model with the largest $Z$ being favoured. According to the scale of \citet{jeffreys1961theory}, to completely rule out a model, an odds ratio of $10^{-2}$ is required. The CDM profile has the largest Bayesian evidence and the Bayes factors indicate that the preference of CDM over SFDM is substantial ($Z < 10^{-0.66}$). However, this value is still far from significant, and we cannot decisively rule it out. 

\subsection{Velocity anisotropy profile}
\begin{figure}[h]
   \centering
   \includegraphics[width=\columnwidth]{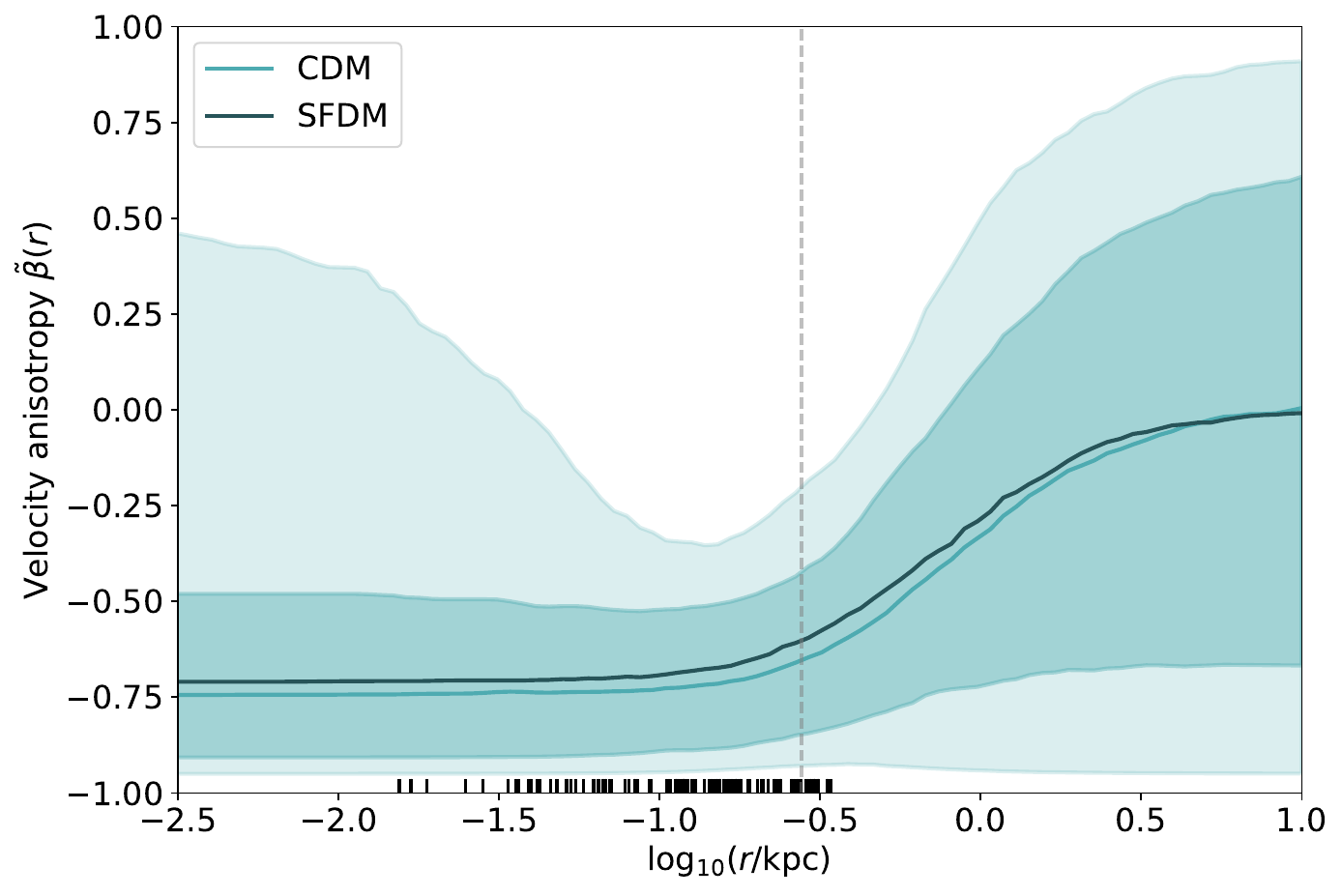}
   \caption{Velocity anisotropy of Ant B. The best fit from \textsc{GravSphere} of the symmetrised anisotropy, $\Tilde{\beta}$, for the CDM model is shown as a solid light blue line, with the light blue shaded regions showing the 68$\%$ and 95$\%$ confidence intervals for the CDM model. The solid dark blue line represents the best fit for the SFDM model. Since the best fit for both models is so similar, the confidence intervals for the SFDM model are omitted for clarity. The half-light radius is represented by the vertical dashed line and the bottom marks represent the projected radii of the members of Ant B.}\label{fig:anis}
\end{figure}

\cite{collins2021} argue that we can use tighter priors on the symmetrised velocity anisotropy, $\Tilde{\beta}(r)$, since dynamical systems in pseudo-equilibrium theoretically should have an isotropic distribution close to the centre, with radial or weak tangential anisotropy at large radii (see e.g. \citealt{read2006}, \citealt{pontzen2015}, \citealt{alvey2021}, and \citealt{orkney2021}). Accordingly, they used $\Tilde{\beta}(r) > -0.1$ and $\Tilde{\beta}(r) \rightarrow 0$ for $r\rightarrow 0$ as default values in the most recent version of \textsc{GravSphere}. However, when we first used these priors on the anisotropy profile, the values obtained for the density and mass profile were extremely low. For this reason, we decided to allow the full range of values on $\Tilde{\beta}(r)$. We then found out that Ant B prefers a negative anisotropy over all radii, although it becomes more positive for larger radii. This behaviour can be seen in Figure~\ref{fig:anis}. Since all models followed a similar behaviour in the anisotropy profile, the CDM profile was chosen for display, for it is the standard model and has proven to be representative of all tested models. The corner plots of the relevant anisotropy parameters for both models are shown in Figure~\ref{fig:corner_beta} of the Appendix. \par
Given the preference for a tangential anisotropy, we also ran the models with tighter priors on the anisotropy parameters, but this time allowed only for negative values, with $-0.6 < \Tilde{\beta}_0 < 0.4$ and $ -0.6 < \Tilde{\beta}_\infty < -0.4$. We get reasonable density profiles for both models but the velocity dispersion does not fit properly, which suggests that Ant B does not have a constant anisotropy profile, and therefore we need to allow for the full range of these parameters.

The anisotropy profile of this dwarf favours tangential anisotropy at its centre and becomes isotropic - and possibly radial - after the half-right radius. As mentioned, this behaviour is not expected for a system like Ant B, since it is thought that this galaxy is in pseudo-equilibrium. This feature could be due to several reasons, ranging from tidal interaction with the host galaxy, a gas disk formed before the stars, unmodelled rotation, or even the presence of a star cluster contaminating the inner bin. We will discuss the astrophysical implications and investigate the origin of this result in a companion paper (Brinchmann et al., in prep.). The impact that this feature has on the dark matter constraints will be discussed in Sect.~\ref{subsect:robustness}. 


\section{Discussion} \label{discussion}
The virial mass, $M_{200} \sim 10^9$M$_\odot$, and the concentration parameter, $c_{200}$, found for this system using both the CDM and the SFDM models are consistent with it being a gas-rich ultra-faint like Leo T \citep{vaz}. The implications of our findings for galaxy formation models will also be discussed in detail in Brinchmann et al., in prep; see also \citealt{bas2022}. \par 

\subsection{Constraints on the dark matter models}
Turning now to the dark matter constraints that we were able to obtain, to match the observations of dwarf galaxies in the Local Group that show cores of the size $~\sim 1$ kpc, the SI strength parameter has to satisfy $R_{\text{TF}} \sim 1$ kpc $\Longleftrightarrow \frac{g}{m^2c^4}\lesssim 2 \times 10^{-18}$ eV$^{-1}$cm$^3$. Furthermore, \citet{sfdm2014, sfdm2017} were able to place upper limits on the values of $m$ and $g$ based on observational consequences while assuming that dark matter is SFDM. They took into account all the phases in the evolution of the Universe and how the value of $g/m^2$ would influence the duration of these different phases. The redshift required by the observations of the CMB temperature anisotropy power spectrum for the transition from a radiation-dominated Universe to a matter-dominated one places an upper limit for $g/m^2$ of $\frac{g}{m^2c^4}\lesssim 4 \times 10^{-17}$ eV$^{-1}$cm$^3$, corresponding to $R_{\text{TF}} \leq 5$ kpc. Additionally, \citet{sfdm2021} were able to get an estimate of the minimum particle mass to ensure that for the SFDM-TF regime $R_{\text{TF}} \gg \lambda_{\text{deB}}$, given by
\begin{equation}
    \frac{mc^2}{10^{-21}\text{ eV}} \gg \left(\frac{M_{200,\text{min}}}{10^9\text{ M}_\odot}\right)^{-1/3}\left(\frac{R_{\text{TF}}}{1\text{ kpc}}\right)^{-1},
    \label{sfdmparticlemass}
\end{equation}
where $M_{200,\text{min}}$ is the minimum halo mass that needs to be accommodated by the model. \par
Our upper limit on the characteristic length scale is $R_{\text{TF}} \lesssim 0.18$ kpc at the $68\%$ confidence level and $R_{\text{TF}} \lesssim 0.72$ kpc at the $95\%$ confidence level, which translates to $\frac{g}{m^2c^4}\lesssim 5.2 \times 10^{-20}$ eV$^{-1}$cm$^3$ at the $68\%$ confidence level and $\frac{g}{m^2c^4}\lesssim 8.3 \times 10^{-20}$ eV$^{-1}$cm$^3$ at the $95\%$ confidence level. Using these values we can also determine the minimum particle mass $m \gg 4.8 \times 10^{-21}$ eV$/c^2$. 
To give a visual illustration of the tension between the $R_{\text{TF}}$ required by \citet{sfdm2021} and what we find, Figure~\ref{fig:dens} also contains a density profile for Ant B with $R_{\text{TF}}=1$ kpc, assuming that $M_{200}$ and $c_{200}$ have the median values found for SFDM above. As is clear, the required density profile is strongly inconsistent with the one we find. In Figure~\ref{fig:veldisp_recovered} we also show the shape that the velocity dispersion of Ant B would have if $R_{\text{TF}}=1$ kpc, assuming that the free parameters of the SFDM have the same median values previously found. \par
The lack of line-of-sight velocities for stars very close to the centre limits our ability to constrain the density profile below 30 pc (check Figure~\ref{fig:dens}), and hence our ability to place stronger constraints on SFDM. \par
More recently, \cite{shapiro} revisited structure formation in the SFDM in the cosmological context to understand if the requirement of $R_{\text{TF}} \gtrsim 1$ kpc was consistent with the cosmological formation of these haloes. To do this, they applied the equations previously derived in \cite{sfdm2021} to simulate individual halo formation by spherical infall and collapse in a cosmologically expanding universe. Using CDM-like initial conditions, they found that the density profiles were similar to those found using non-cosmological initial conditions, being consistent with the former requirement of $R_{\text{TF}} \gtrsim 1$ kpc. However, the initial conditions for SFDM halo formation may differ from those for CDM. To overcome this, they used linear perturbation theory to estimate the range of $R_{\text{TF}}$-values that are consistent with observational constraints on the FDM model.
In this case, a core as large as $R_{\text{TF}} \gtrsim 1$ kpc is disfavoured. To match the FDM particle masses $1 \times 10^{-22} \lesssim m$ (eV$c^{-2}$) $\lesssim 30 \times 10^{-22}$, a core of $10 \gtrsim R_{\text{TF}}$ (pc) $\gtrsim 1$ is required, favouring sub-kpc core sizes. However, in this range, SFDM approaches CDM, and the problems associated with the standard model start to arise once again. This suggests that either the observational constraints of FDM cannot be used to place the corresponding constraints on the core sizes of the haloes formed in SFDM, or SFDM is an incompatible explanation for the observed cores in the dwarfs of the Local Group. In \cite{hartman2022} and \cite{sfdm_simulations}, similar conclusions were drawn, using large-scale observables and fully 3D cosmological simulations, respectively.

\subsection{Testing the robustness of the results}\label{subsect:robustness}
The centre measured by \cite{sand2015} might be miscalculated since there is a bright star close to it, partially overlapping the galaxy. To test the robustness of the results, we started by changing the centre of Ant B by ($\alpha \sim 1'', \delta \sim 3''$) to see how much our results would be affected. We reran \textsc{Binulator} and \textsc{GravSphere} on both models for Ant B with the new centre. The velocity dispersion for each bin, the recovered velocity dispersion profile, and the anisotropy profile associated with the new centre can be seen in Figures~\ref{fig:veldisp_bin_offset},~\ref{fig:veldisp_offset}, and~\ref{fig:anis_offset} of the Appendix, respectively. Figures~\ref{fig:cornerCDM_offset} and \ref{fig:cornersfdm_offset} show the new parameters derived for these profiles. We find a virial mass of ${M_{200}\approx 1.29^{+2.02}_{-0.71}\times 10^9}$ M$_\odot$ and a concentration parameter of ${c_{200} \approx 18.20^{+5.79}_{-4.71}}$ for the CDM model and ${M_{200} \approx 1.35^{+2.04}_{-0.76}\times 10^9}$ M$_\odot$ and  ${c_{200} \approx 17.78^{+5.66}_{-4.60}}$ for the SFDM model. We find ${R_{\text{TF}} \lesssim 0.11}$ kpc at the $68\%$ confidence level and $R_{\text{TF}}\lesssim 0.68$ kpc at the $95\%$ confidence level. Let us notice that this time we ran the code with half the steps to make the process faster since we wanted solely to check whether our results were not dramatically affected by changing the centre. These results are very close to those previously obtained. This is as expected, since the innermost bin after radial binning still contains 20 of the 25 stars that we have with the nominal centre. \par

We find that Ant B favours tangential inner velocity anisotropy for both the CDM and SFDM models. We will discuss the interpretation of these results and the implications for galaxy formation models in a companion paper (Brinchmann et al., in prep.). Since the anisotropy has its lowest value in the centre of the galaxy, we need to test whether the constraints on the dark matter parameters are affected when we remove the stars that are, in principle, responsible for this behaviour. To do this, we removed the 25 stars closest to the centre that correspond to the first bin (represented in Figure~\ref{fig:veldisp_recovered}). We reran \textsc{Binulator} and \textsc{GravSphere} on both models for Ant B without these 25 stars. The velocity dispersion for each bin, the recovered velocity dispersion profile, and the anisotropy profile associated with the new centre can be seen in Figures~\ref{fig:veldisp_bin_removed},~\ref{fig:veldisp_removed}, and~\ref{fig:anis_removed} of the Appendix, respectively. \par 
As can be seen in Figure~\ref{fig:cornerCDM_1stbin}, the parameters of the CDM model suffer almost no changes: we find a virial mass of ${M_{200} \approx 1.55^{+3.24}_{-0.98}\times 10^9}$ M$_\odot$ and a concentration parameter of ${c_{200} \approx 17.38^{+6.06}_{-4.50}}$. In Figure~\ref{fig:cornersfdm_1stbin} it is possible to see how the dark matter constraints for SFDM changed with the removal of the innermost bin. Looking at the physical constraints on $R_{\text{TF}}$, we find a lower value at the 68$\%$ confidence level ($R_{\text{TF}} \lesssim 0.11$ kpc instead of $\lesssim 0.18$ kpc) and a slightly higher value at the 95$\%$ confidence level ($R_{\text{TF}} \lesssim 0.81$ kpc instead of $\lesssim 0.72$ kpc). Similarly to what was done in the previous robustness test, we ran the code with half the steps. \par 
The parameters associated with the anisotropy profile continue to show the same behaviour as before: the anisotropy is tangential in the centre of the galaxy and becomes almost isotropic when we approach larger radii (Figure~\ref{fig:anis_removed}); however, this tangential anisotropy becomes less statistically significant by removing the innermost bin (Figure~\ref{fig:corner_beta_removed}). Since we find that the recovered SFDM model parameters do not change significantly, we consider our SFDM profile results to be robust.


\section{Conclusions}\label{conclusions}
In this work, we present the first spectroscopic observations of Ant B, a distant dwarf found by \citet{sand2015}. With them, we were able to determine kinematic data, previously presented by \citet{bas2022}, and ensure the membership of 127 stars. These observations allowed us to determine the intrinsic mean line-of-sight velocity, $v_{\text{LOS}}=375.39^{+0.98}_{-0.97}$ km s$^{-1}$, and the velocity dispersion, $\sigma_{v_{\text{LOS}}}=7.87^{+1.02}_{-0.98}$ km s$^{-1}$, of this galaxy. The estimated $v_{\text{LOS}}$ is in agreement with the velocity of \hi\ gas present in this galaxy ($v_{\hi} \sim 375$ km s$^{-1}$). \par 
The kinematic data of Ant B also allowed us to derive the dark matter density profiles for this dwarf. From them we were able to determine the virial mass, $M_{200} \approx 10^{9.2}$M$_\odot$, and the concentration parameter, $c_{200} \approx 17$, of Ant B. The values found for these parameters for both dark matter models are consistent with the values expected from models in which the smallest dwarf galaxies are reionization fossils. \par 
We find that Ant B favours tangential anisotropy in its centre, although it becomes more positive for larger radii, showing an isotropic behaviour - and possibly radial anisotropy - after the half-right radius. \par
We find substantial Bayesian evidence ($Z < 10^{-0.66}$) against the SFDM model. However, it is not significant, and we cannot completely rule out this model based solely on this result. \par
We constrained the characteristic length scale of the repulsive SI $R_{\text{TF}}$ of the SFDM model of $R_{\text{TF}} \lesssim 180$ pc ($68\%$ confidence level) and $R_{\text{TF}} \lesssim 720$ pc ($95\%$ confidence level), which translates to $\frac{g}{m^2c^4}\lesssim 5.2 \times 10^{-20}$ eV$^{-1}$cm$^3$ ($68\%$ confidence level) and $\frac{g}{m^2c^4}\lesssim 8.3 \times 10^{-20}$ eV$^{-1}$cm$^3$ ($95\%$ confidence level). This gives us a minimum particle mass of $m \gg 4.8 \times 10^{-21}$ eV$/c^2$. Though we cannot rule SFDM entirely, we find a constraint on $R_{\text{TF}}$ that rules out it being a solution to the core-cusp problem. \par 
Tests show that the derived constraints are robust against different model assumptions, such as changing the location of the centre or removing the stars closest to it, with minimal impact on the final results. \par
The $R_{\text{TF}}$ that we find is not large enough to solve the core-cusp problem and therefore this model cannot explain cores in dwarfs in the Local Group. \par 
To further improve the constraints placed on the SFDM model, the line-of-sight velocities of stars closer to the centre would be required. This remains challenging since not only is Ant B a distant and faint galaxy, but it also has a bright foreground star partially overlapping it. \par 
The dark matter density profiles of UFDs have only started being estimated very recently. The measurements of the velocities of stars in these galaxies with the spectroscopic data from MUSE-Faint allow us to improve the constraints on their inner dark matter density profiles. With a larger number of UFDs analysed, we will be able to place stronger constraints on the physical properties of dark matter. Furthermore, we would also like to explore alternative dark matter models, which will take us one step closer to understanding this unknown form of matter.

\begin{acknowledgement}
We thank the anonymous referee for their helpful comments, which improved the manuscript. MPJ thanks Marcel S. Pawlowski, Pengfei Li, Salvatore Taibi and Jamie K. Kanehisa for the valuable discussions and advice. MPJ and JB acknowledge support by Fundação para a Ciência e a Tecnologia (FCT) through the research grants UID/FIS/04434/2019, UIDB/04434/2020, UIDP/04434/2020 and PTDC/FIS-AST/4862/2020. JB acknowledges support from FCT work contract 2020.03379.CEECIND.
SLZ acknowledges support by The Netherlands Organisation for Scientific Research (NWO) through a TOP Grant Module 1 under project number 614.001.652. SKA acknowledges funding from UKRI in the form of a Future Leaders Fellowship (grant number MR/T022868/1). Based on observations made with ESO Telescopes at the La Silla Paranal Observatory under programme IDs 0100.D-0807, 0101.D-0300, 0102.D-0372 and 0103.D-0705. This research has made use of Astropy \citep{astropy}, corner.py \citep{corner}, matplotlib \citep{matplotlib}, NASA’s Astrophysics Data System Bibliographic Services, NumPy \citep{numpy}, SciPy \citep{scipy}.  Based on observations made with the NASA/ESA Hubble Space Telescope, and obtained from the Hubble Legacy Archive, which is a collaboration between the Space Telescope Science Institute (STScI/NASA), the Space Telescope European Coordinating Facility (ST-ECF/ESA) and the Canadian Astronomy Data Centre (CADC/NRC/CSA). 
\end{acknowledgement}

\bibliographystyle{aa}
\bibliography{Constraining-SFDM-AntliaB}

\begin{appendix}
\section{Numerical parametrization of SFDM}\label{app:parametrization}
The estimation of the SFDM model is not trivial since, to determine $\alpha$ and $\rho_c$ for a given halo to obtain the density profile, one has to ensure continuity in the density profile and mass conservation at the switch point~\citep{sfdm2021} described, respectively, by
\begin{equation}
\rho_c\sinc(\pi\alpha) = \rho_{\text{CDM}}(\alpha R_{\text{TF}})
\label{eq:continuity}
\end{equation}
and
\begin{equation}
\int^{\alpha R_{\text{TF}}}_0 \rho_c\sinc\left(\frac{\pi r}{R_{\text{TF}}}\right)4\pi r^2\dd r =\int^{\alpha R_{\text{TF}}}_0 \rho_{\text{CDM}}(r)4\pi r^2\dd r.
\label{eq:massconservation}
\end{equation}
Integrating both sides of Equation~\ref{eq:massconservation}, we get
\begin{multline}
    \frac{4}{\pi^2}R_{\text{TF}}^3\rho_c(\pi\alpha\cos(\pi\alpha)-\sin(\pi\alpha)) \\ +
    4\pi\rho_0 r_s^3\left[\ln\left(\frac{r_s+\alpha R_{\text{TF}}}{r_s}\right)+\frac{r_s}{r_s+\alpha R_{\text{TF}}}-1\right] = 0.
    \label{eq:massconsint}
\end{multline}
Now, we can uncouple the unknown parameters $\alpha$ and $\rho_c$. Equation~\ref{eq:continuity} can be written as
\begin{equation}
\rho_c = \frac{\pi\alpha}{\sin(\pi\alpha)}\cdot\frac{\rho_0}{\alpha R_{\text{TF}}/r_s(1+\alpha R_{\text{TF}}/r_s)^2}.
\label{eq:rhoc}
\end{equation}
Thus, Equation~\ref{eq:massconsint} loses its dependence in $\rho_c$, allowing us to solve it for $\alpha$
\begin{multline}
\pi^2r_s^3\left[\ln\left(\frac{r_s+\alpha R_{\text{TF}}}{r_s}\right)+\frac{r_s}{r_s+\alpha R_{\text{TF}}}-1\right] \\ -\alpha R_{\text{TF}}^3\cdot\frac{1-\pi\alpha\cot(\pi\alpha)}{\alpha R_{\text{TF}}(1+\alpha R_{\text{TF}}/r_s)^2} = 0.
\label{eq:alpha}
\end{multline}
The solution of Equation~\ref{eq:alpha} can be found using a root solver and then $\rho_c$ can be easily computed. However, this equation admits two solutions and one of them leads to a negative $\rho_c$, which is not a physical solution. For this reason, one has to carefully choose the interval at which $\alpha$ leads to positive values of $\rho_c$.
After several tries and an extensive study of this equation, we find that a reasonable range to look for a solution is $0.4 < \alpha < 0.99$. With both $\alpha$ and $\rho_c$ determined, the mass profile can be determined:
\begin{equation}
M_{\text{SFDM}}(r) =
\begin{cases}
     4\pi\rho_c\displaystyle{\int_0^r} r^2\sinc(r/R_{\text{TF}}) \dd r & r \leq \alpha R_{\text{TF}} \\
     M_{\text{CDM}} & r \geq \alpha R_{\text{TF}}
    \end{cases}.
    \label{msfdm}
\end{equation}
The density profile is now also easily computed through Equation~\ref{eq:sfdm}.

\begin{figure}[H]
   \centering 
   \includegraphics[scale=0.42]{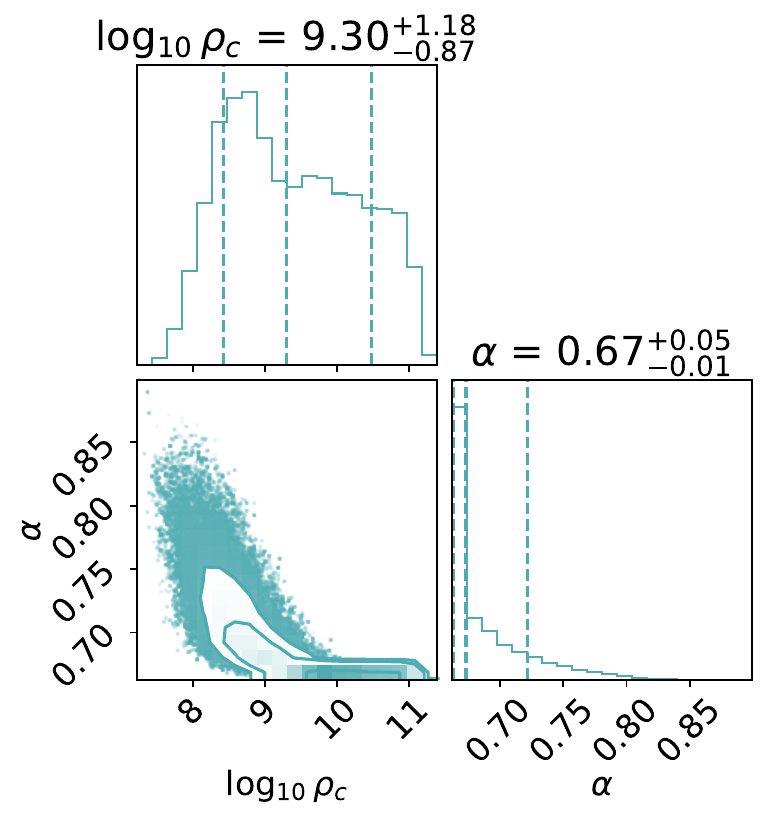}
   \caption[Corner plot of the computational parameters $\alpha$ and $\rho_c$, used in the parametrization of the SFDM model.]{Corner plot of the computational parameters $\alpha$ and $\rho_c$ used in the parametrization of the SFDM model.}\label{fig:corner_SFDM_rho}
\end{figure}

\section{\textsc{GravSphere} fits}\label{app:gravspherefits}

\begin{figure}[H]
   \centering
    \includegraphics[width=\columnwidth]{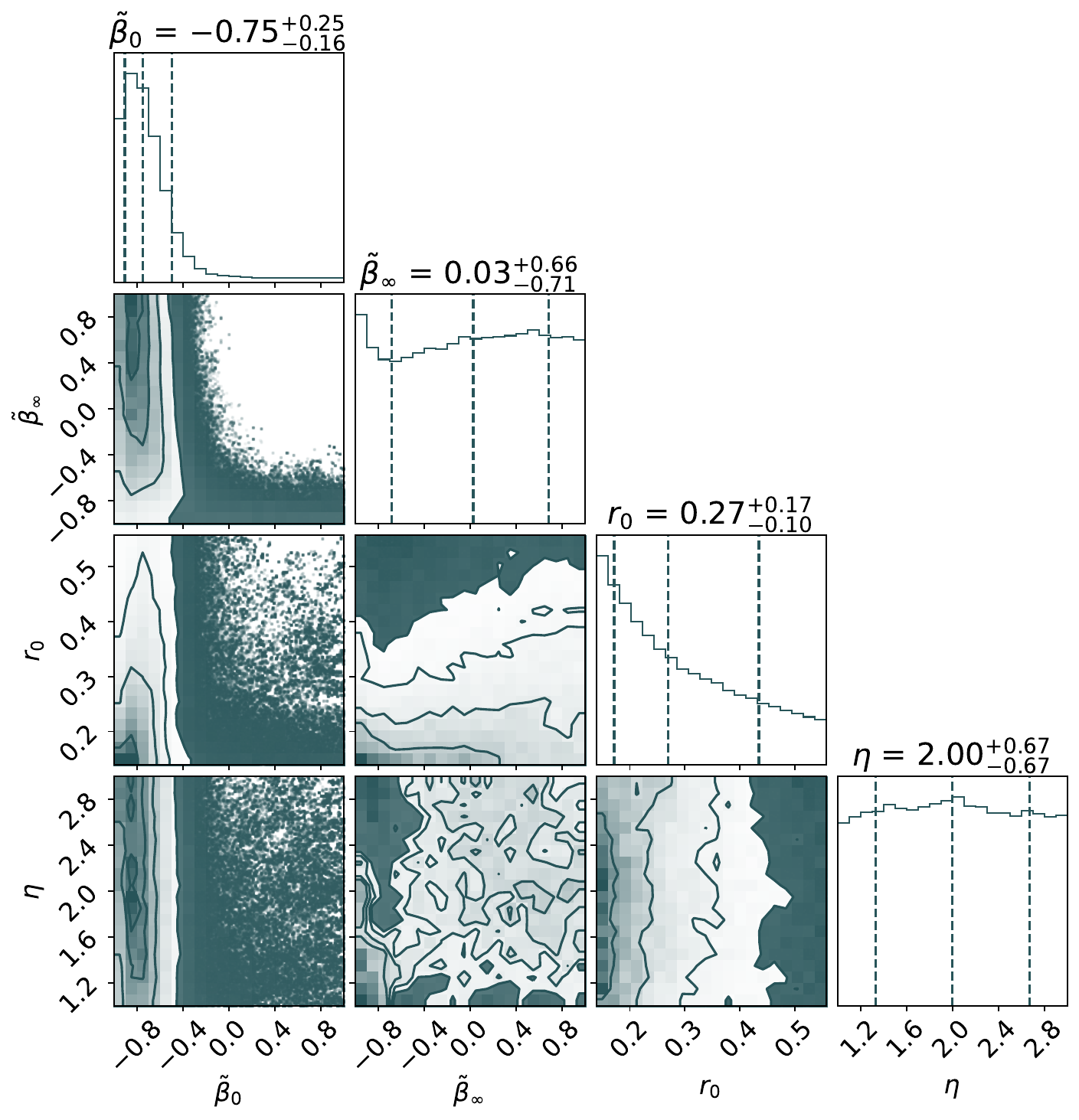} \\%
    \includegraphics[width=\columnwidth]{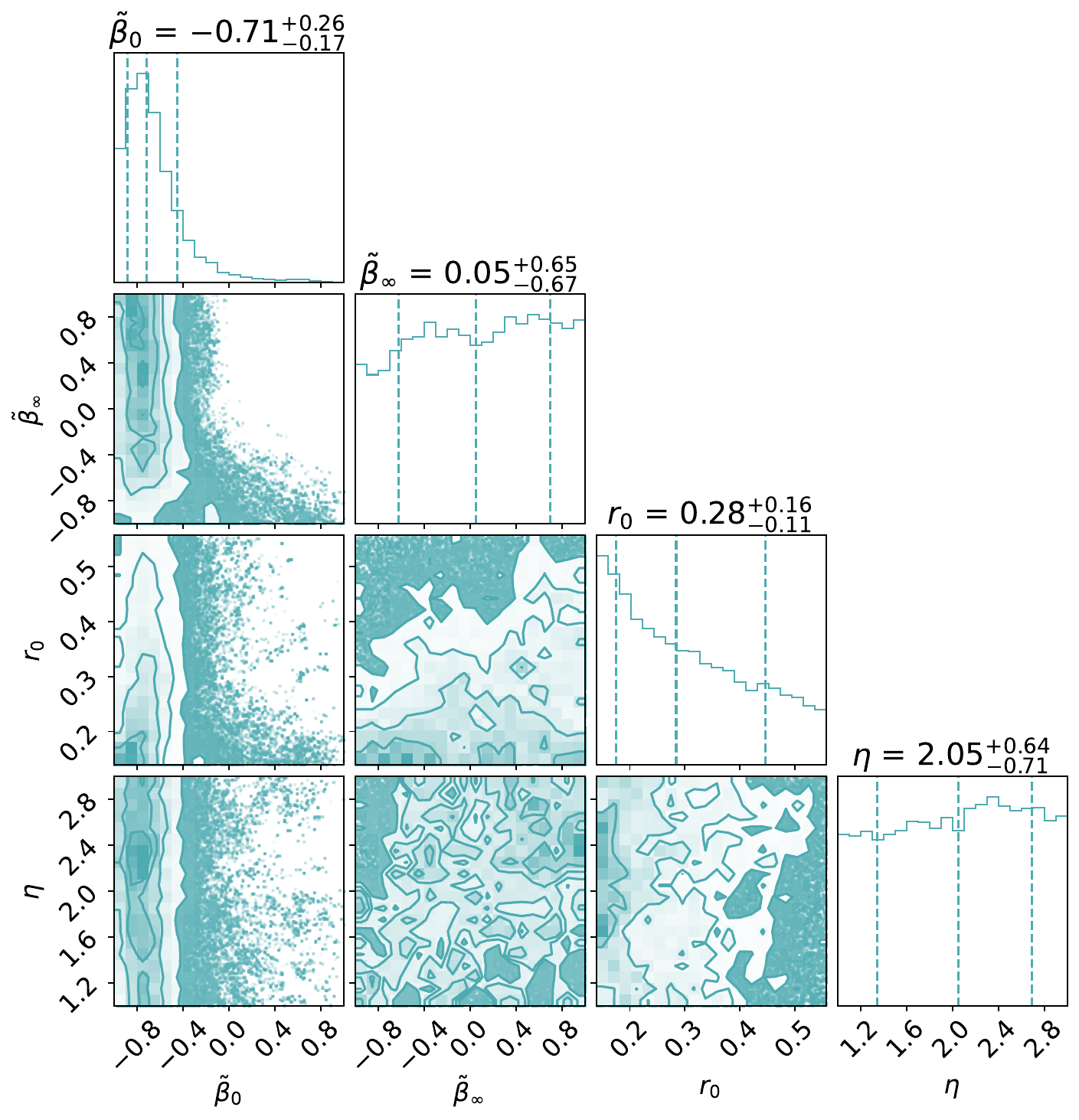}%
    \caption{Corner plot of the parameters $\Tilde{\beta}_0$, $\Tilde{\beta}_\infty$, $r_0$, and $\eta$ of the anisotropy profile for both models. \textbf{Top:} For the CDM model. \textbf{Bottom:} For the SFDM model.}%
    \label{fig:corner_beta}%
\end{figure}

\begin{figure*}
    \centering
    \includegraphics[scale=0.4]{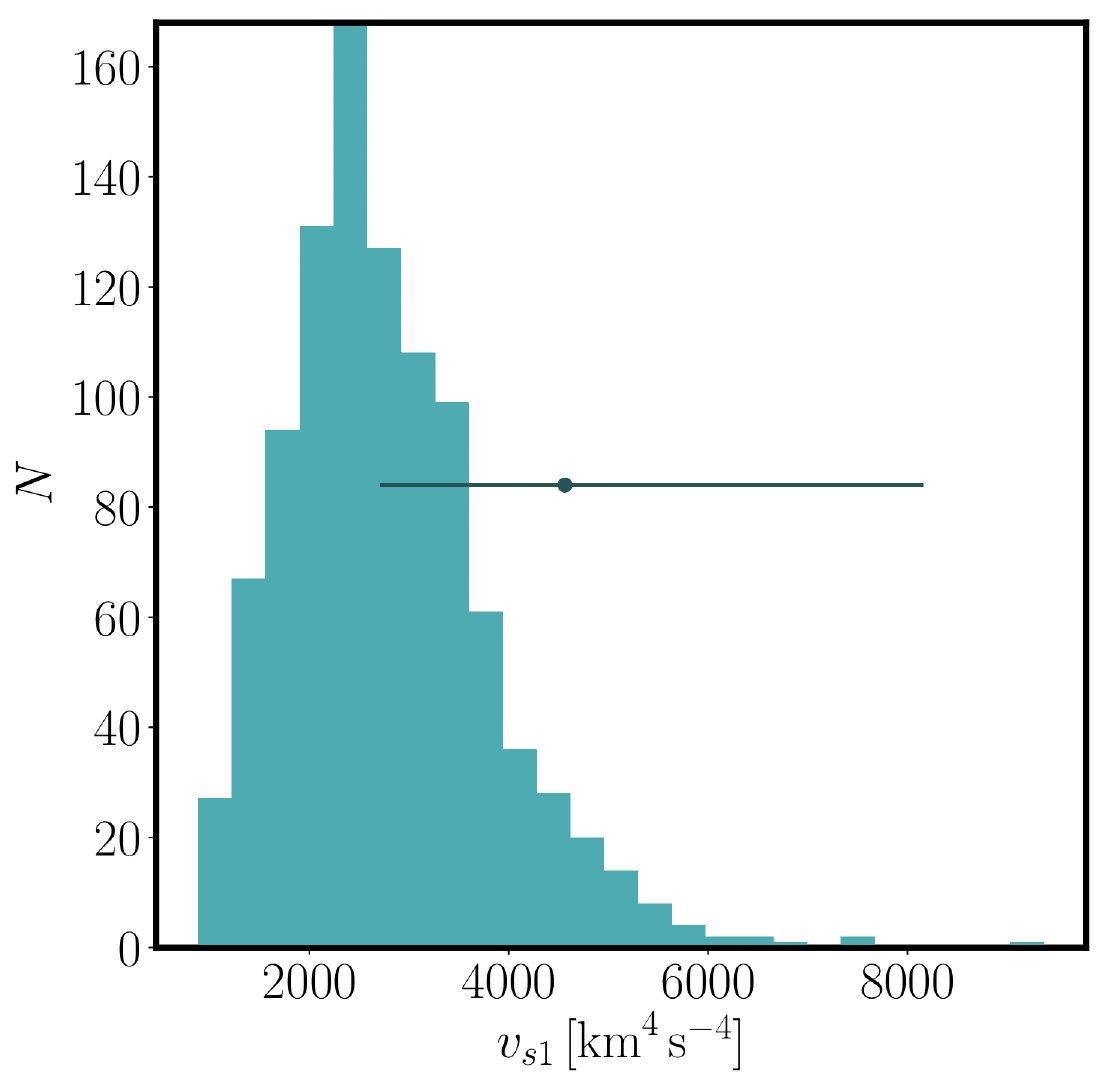}%
    \includegraphics[scale=0.4]{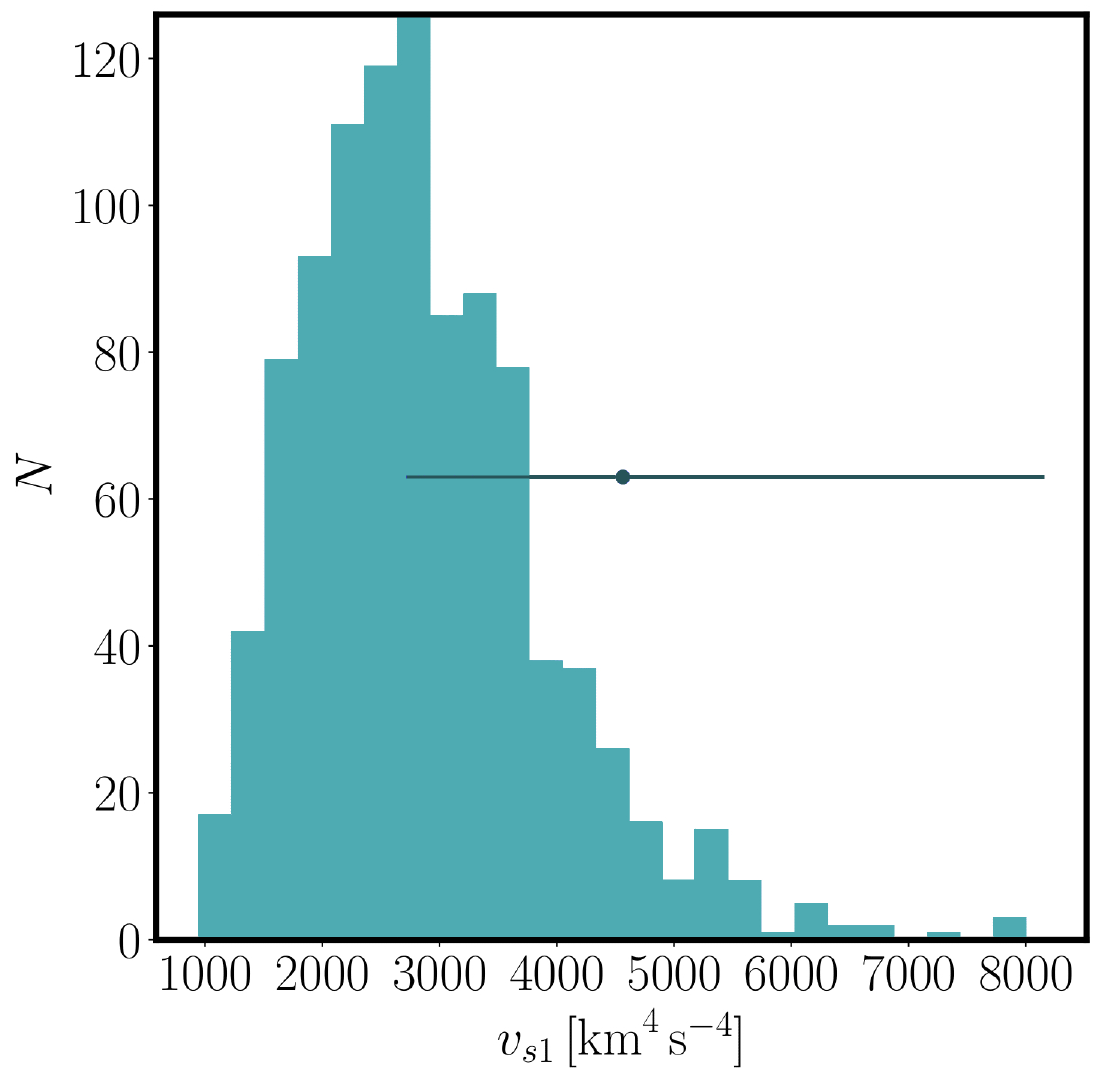} \\
    \includegraphics[scale=0.4]{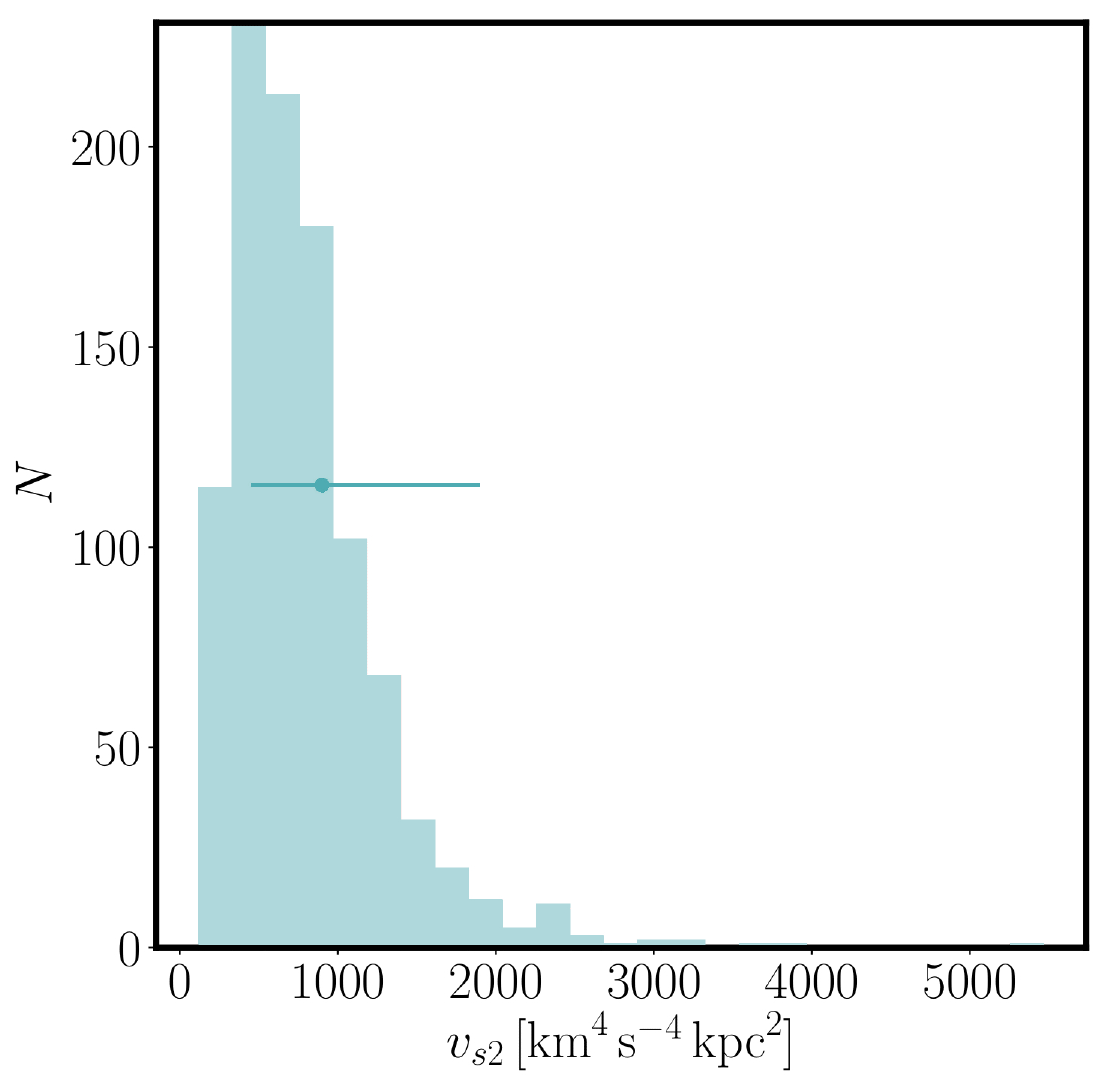}%
    \includegraphics[scale=0.4]{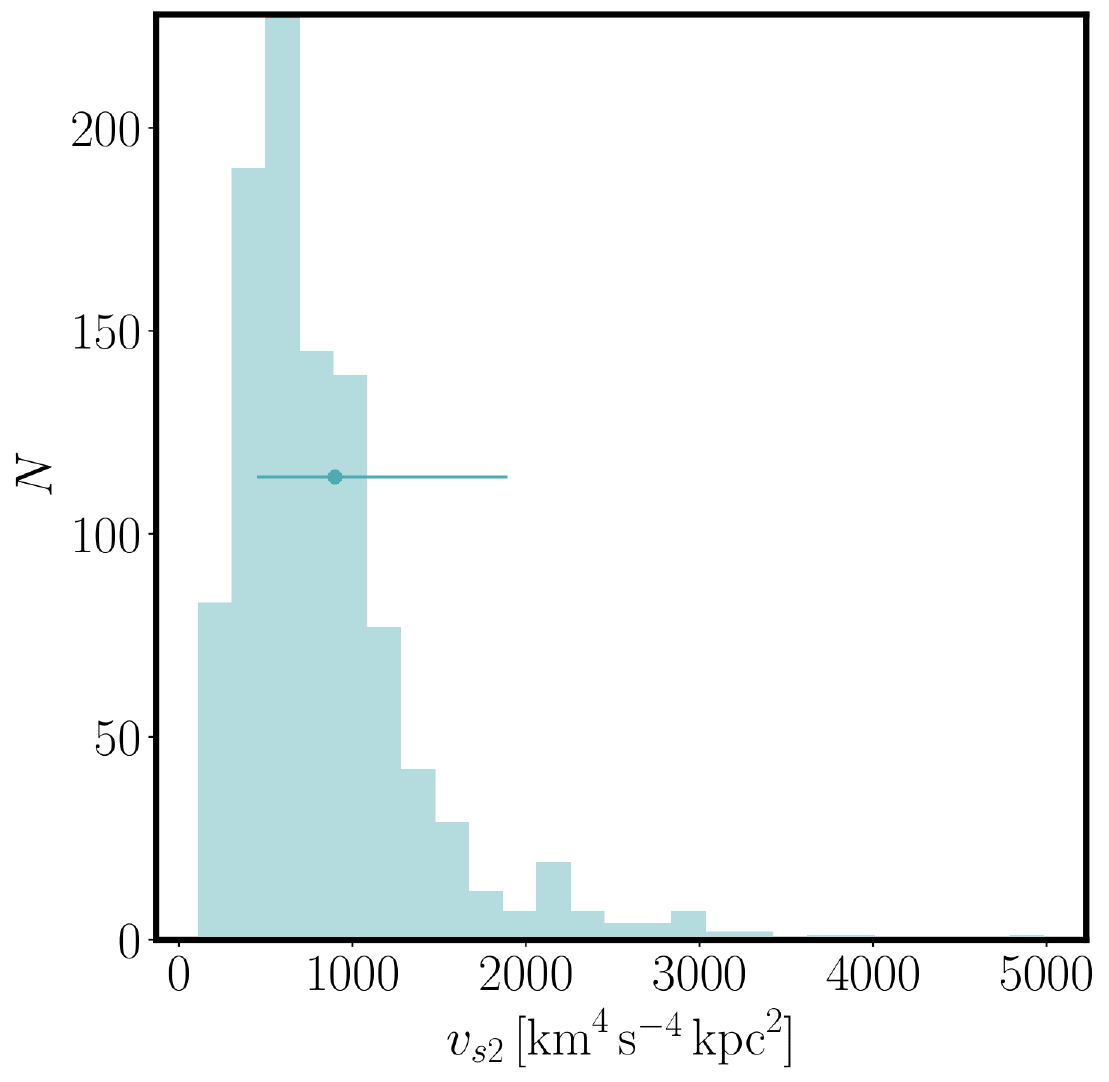} \\
    \includegraphics[scale=0.4]{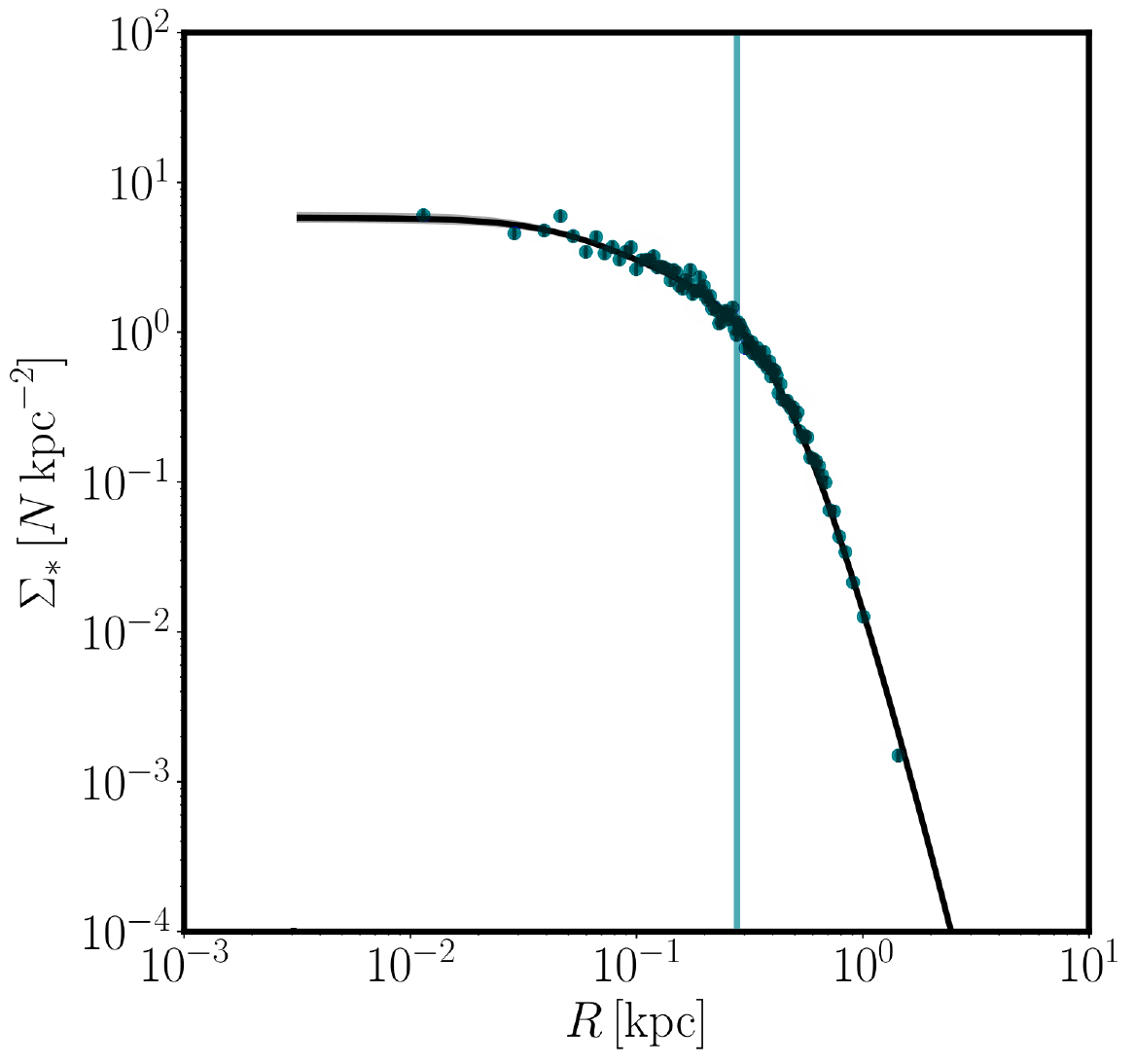}%
    \includegraphics[scale=0.4]{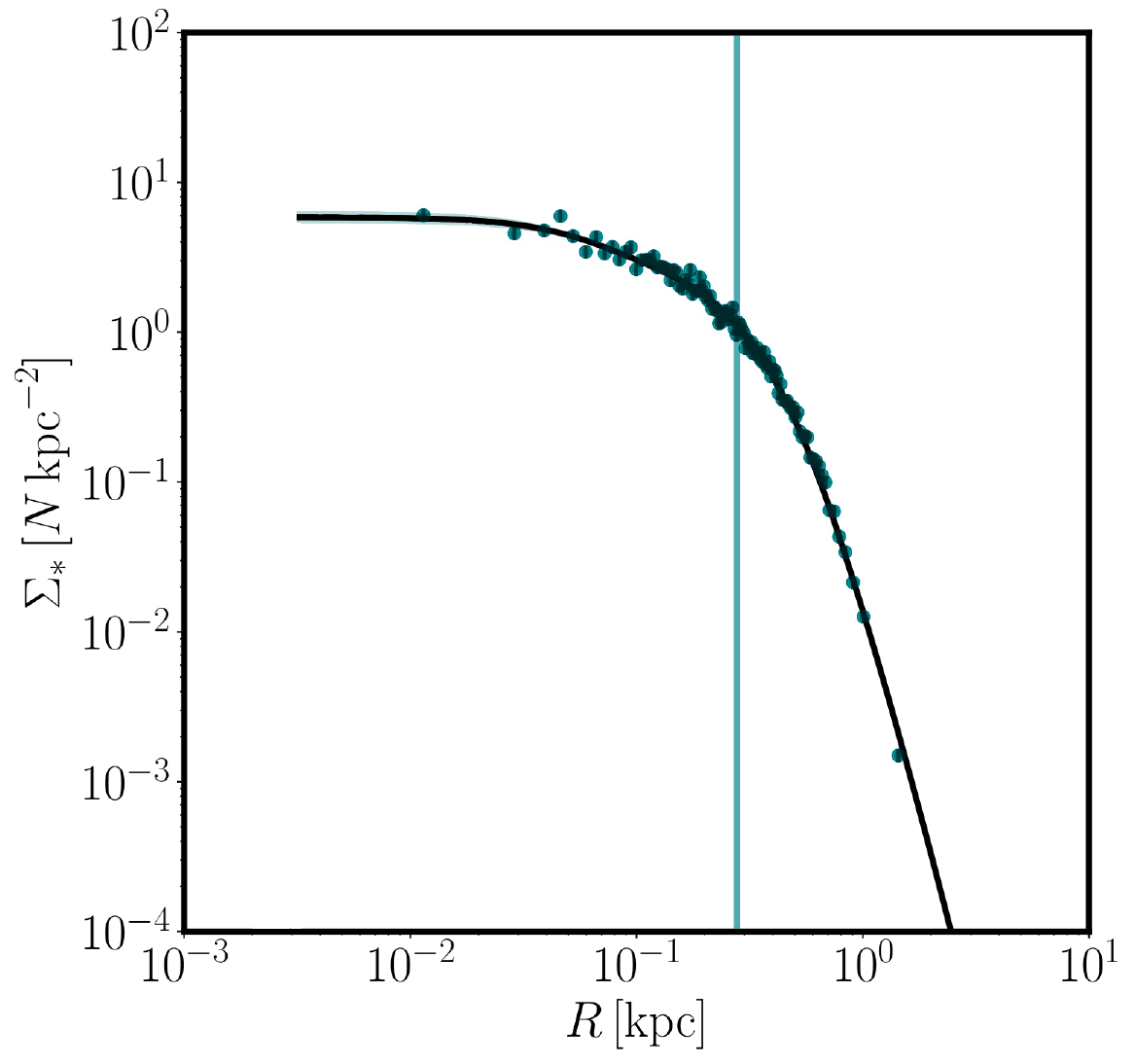}%
      \caption{\textsc{GravSphere} fits for the VSPs and surface brightness profile for both models. \textbf{Top and centre:}  Virial Shape Parameters ($v_{s1}$ on top and $v_{s2}$ in the centre, for both the CDM model (left) and the SFDM one (right). \textbf{Bottom:}  The surface brightness profile, $\Sigma_\star$, for Ant B. The blue points show the photometric data. 
      The best fit from \textsc{GravSphere} is shown as a solid black line, with the shaded grey regions showing the 68$\%$ and 95$\%$ confidence intervals for the CDM model (right) and the SFDM model (left).}%
    \label{fig:vsps_surfb}
\end{figure*}

\FloatBarrier

\section{Robustness of the constraints}\label{app:robustnessplots}
\subsection{Changing the centre}
\begin{figure}[ht]

   \centering
   \includegraphics[width=\columnwidth]{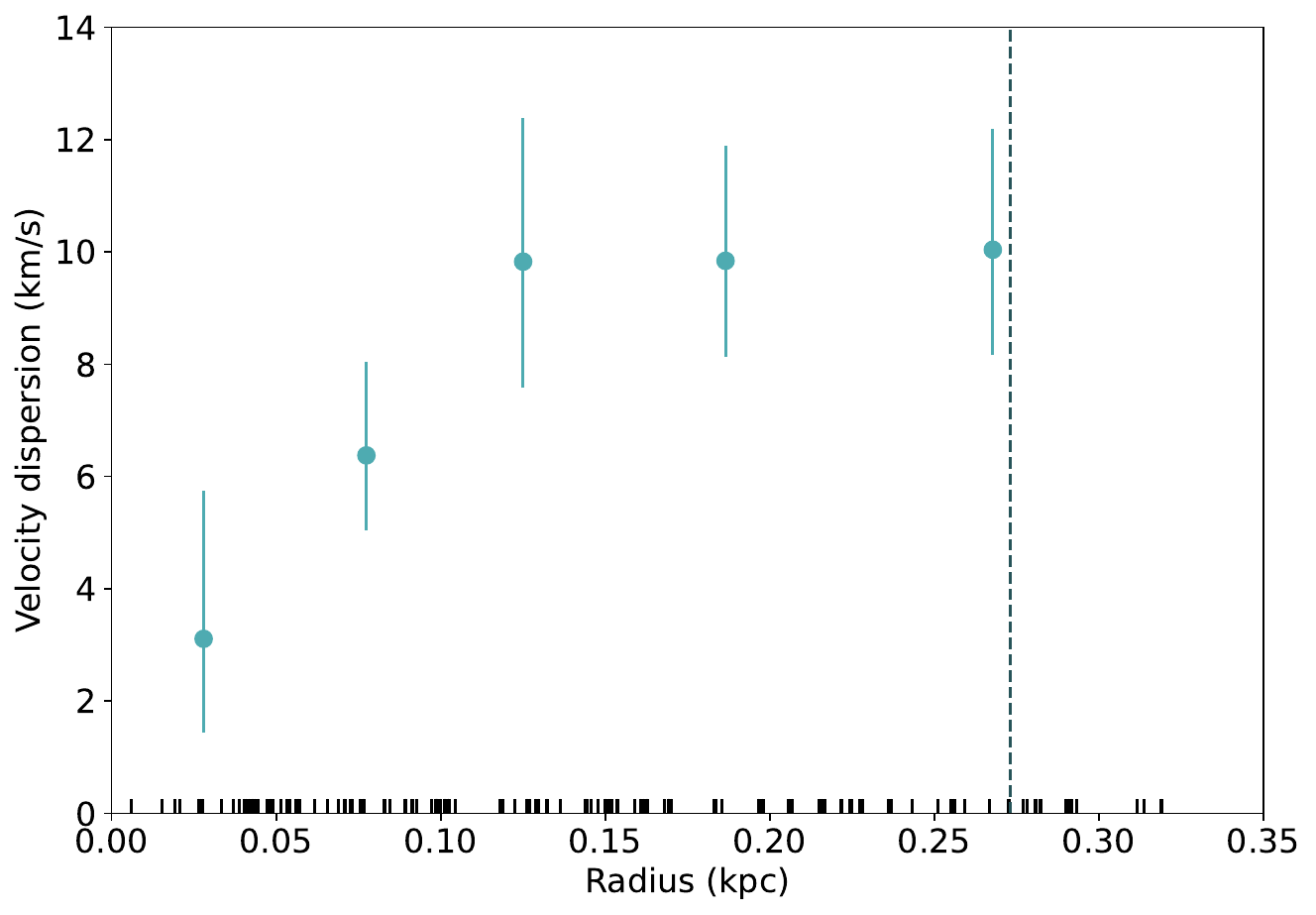}
   \caption[]{Velocity dispersion for each bin along with associated uncertainties after changing the coordinates of the centre of Ant B. The vertical dashed line indicates the half-light radius and the bottom marks represent the projected radii of the members of Ant B.}\label{fig:veldisp_bin_offset}
\end{figure}

\begin{figure}[ht]
   \centering
    \includegraphics[width=\columnwidth]{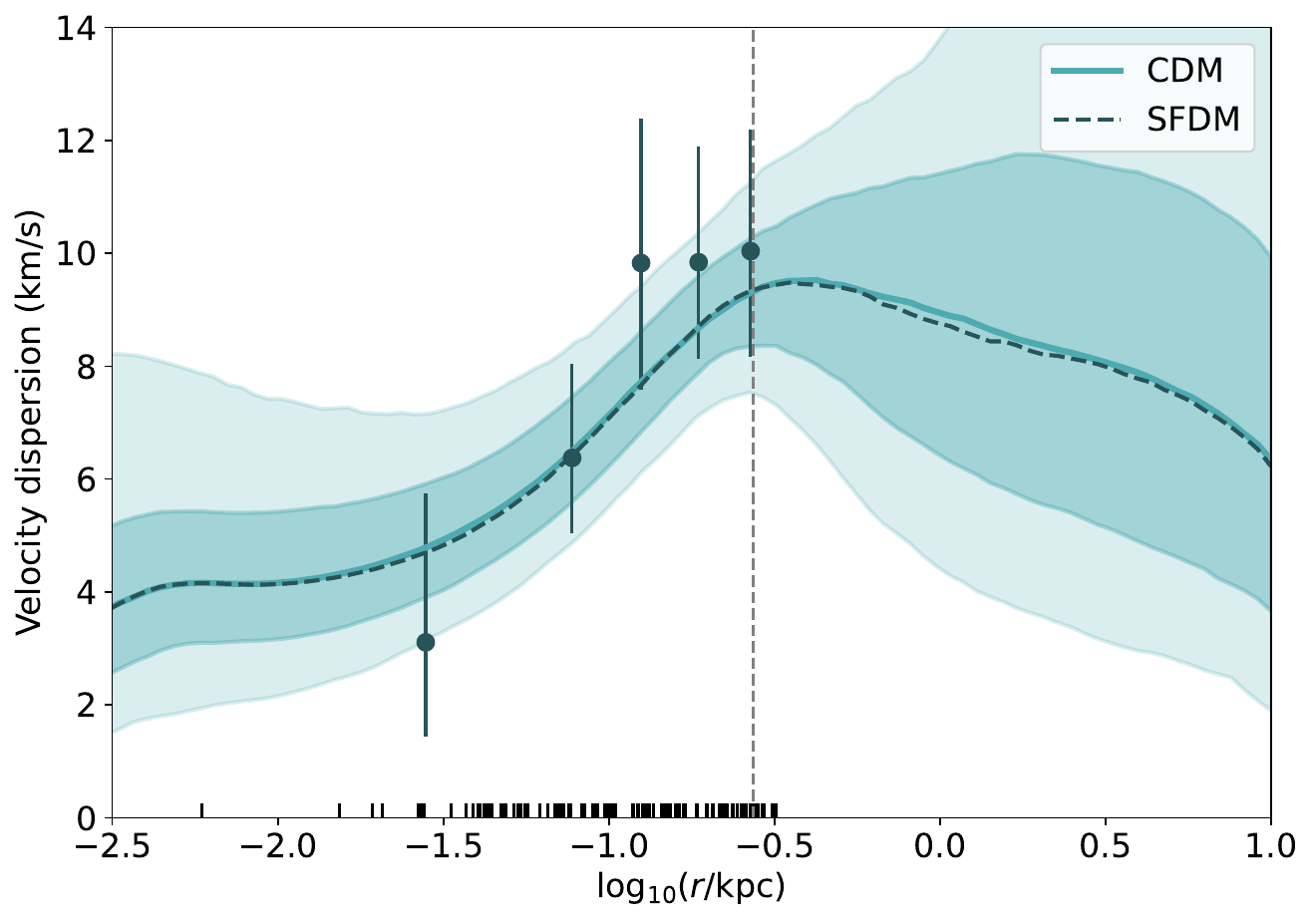}
   \caption{Velocity dispersion of Ant B after changing the coordinates of the centre. The dark blue points represent the binned velocity dispersion previously shown. The best fit from \texttt{GravSphere} for the CDM model is shown as a solid light blue line, with the light blue shaded regions showing the 68$\%$ and 95$\%$ confidence intervals for the CDM model. The dashed dark blue line represents the best fit for the SFDM model. Since the best fit for both models is so similar, the confidence intervals for the SFDM model are omitted for clarity. The half-light radius is represented by the vertical dashed line and the bottom marks represent the projected radii of the members of Ant B.}\label{fig:veldisp_offset}
\end{figure}

\begin{figure}[ht]
   \centering
   \includegraphics[width=\columnwidth]{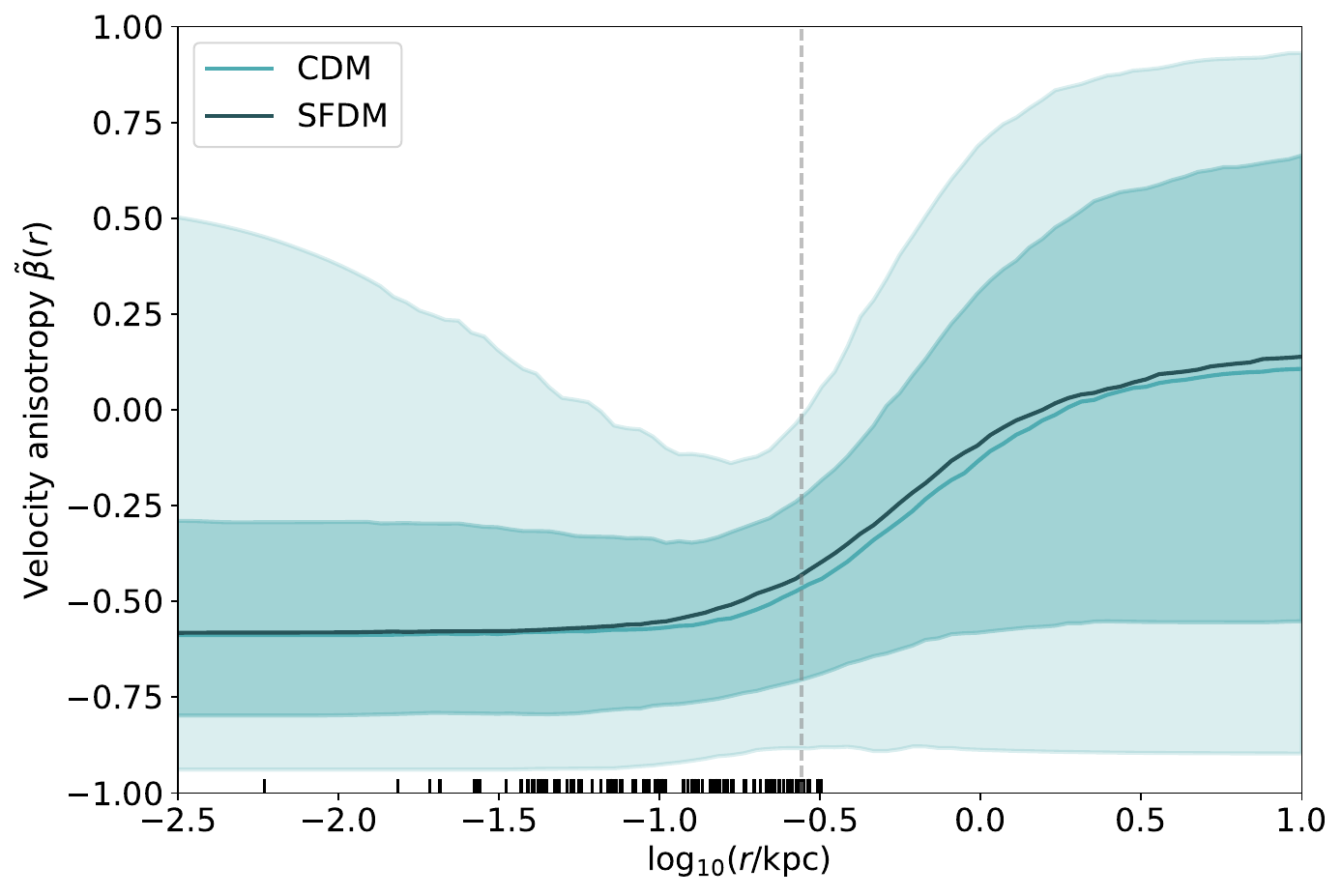}
   \caption{Velocity anisotropy profile of Ant B after changing the coordinates of the centre.
   The best fit from \textsc{GravSphere} of the symmetrised anisotropy, $\Tilde{\beta}$, for the CDM model is shown as a solid light blue line, with the light blue shaded regions showing the 68$\%$ and 95$\%$ confidence intervals for the CDM model. The solid dark blue line represents the best fit for the SFDM model. Since the best fit for both models is so similar, the confidence intervals for the SFDM model are omitted for clarity. The half-light radius is represented by the vertical dashed line and the bottom marks represent the projected radii of the members of Ant B.}\label{fig:anis_offset}
\end{figure}

\begin{figure}[ht]
   \centering
   \includegraphics[width=\columnwidth]{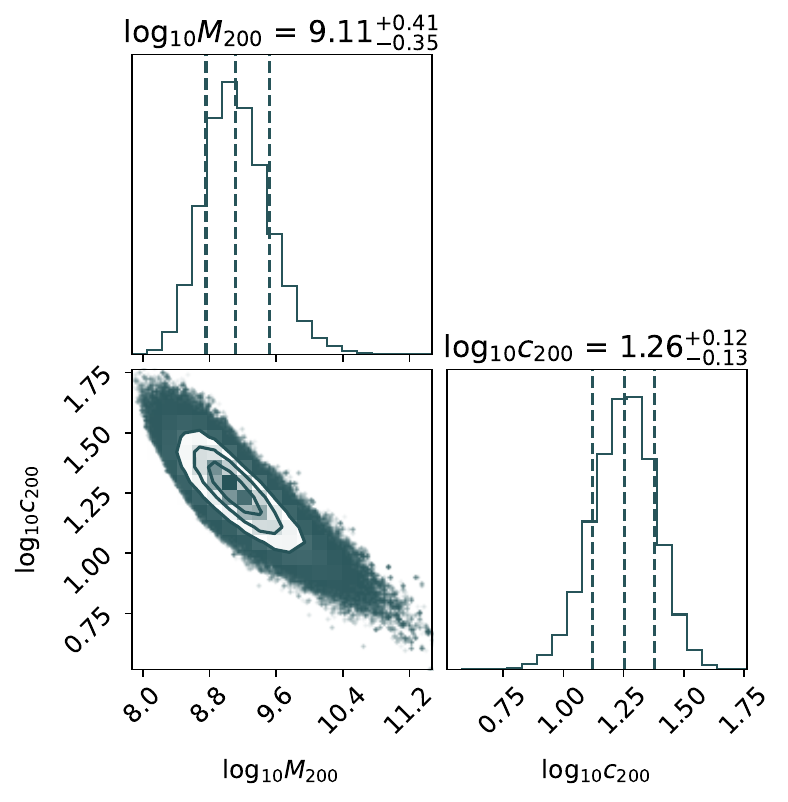}
  \caption{Constraints on the CDM profile for Ant B after changing the coordinates of the centre. The histograms along the diagonal represent the posterior distribution for each parameter: the virial mass, $M_{200}$ in M$_\odot$, and the concentration parameter, $c_{200}$.}\label{fig:cornerCDM_offset}
\end{figure}

\begin{figure}[ht]
   \centering
   \includegraphics[width=\columnwidth]{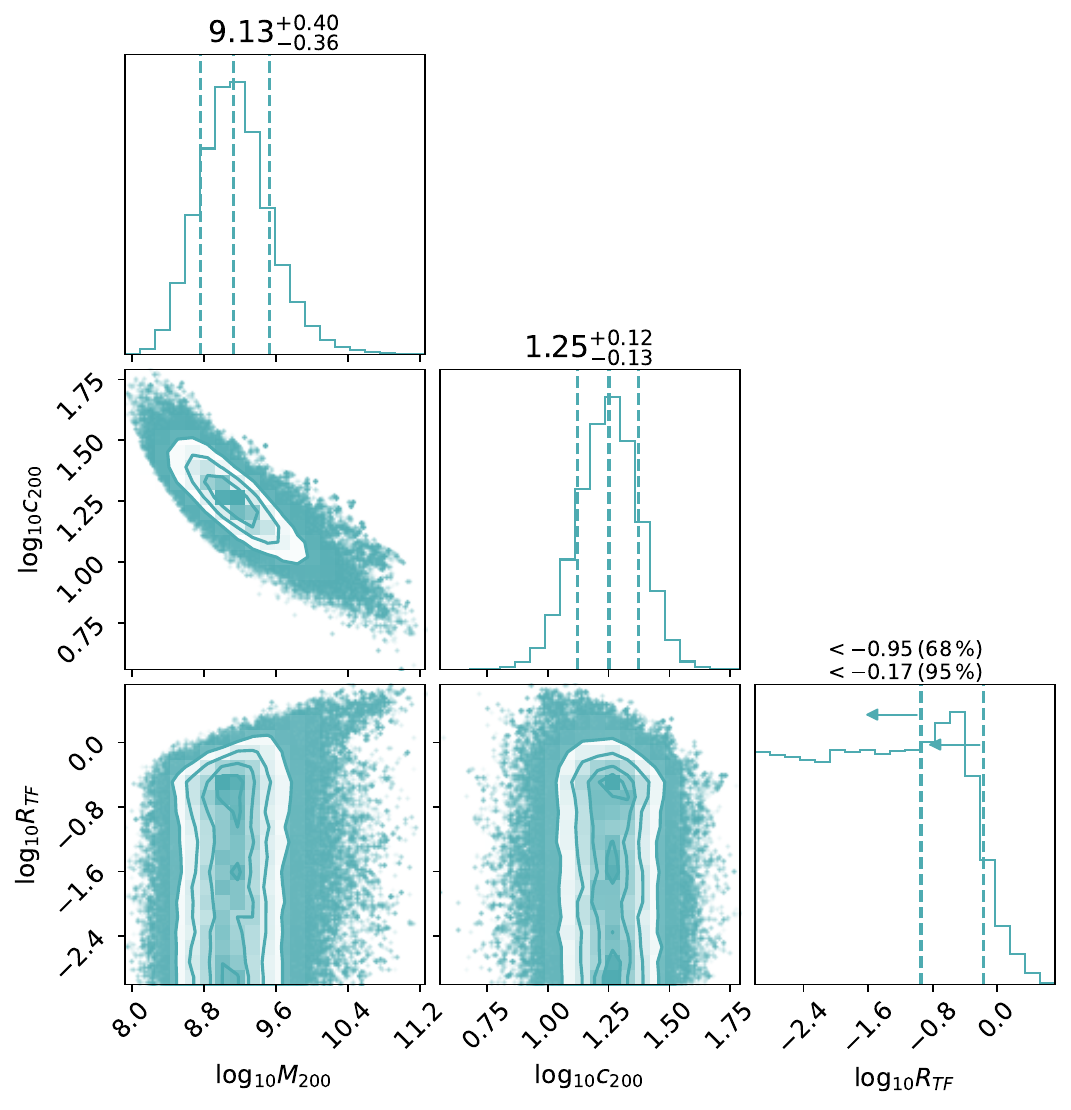}
   \caption{Constraints on the SFDM profile for Ant B after changing the coordinates of the centre. The histograms along the diagonal represent the posterior distribution for each parameter: the virial mass, $M_{200}$ in M$_\odot$, the concentration parameter, $c_{200}$, and the characteristic length scale of the repulsive SI, $R_{\text{TF}}$ in kpc.}\label{fig:cornersfdm_offset}
\end{figure}

The robustness of the constraints of the SFDM model was analysed in two different tests, explained in detail in Sect.~\ref{subsect:robustness}. The first one consisted of changing the centre of Ant B by ($\alpha \sim 1'', \delta \sim 3''$). The velocity dispersion for each bin after this change, the respective recovered velocity dispersion profile, and the anisotropy profile associated with the new centre can be seen in Figures~\ref{fig:veldisp_bin_offset},~\ref{fig:veldisp_offset}, and~\ref{fig:anis_offset}, respectively. The constraints obtained for both the CDM and SFDM profiles for Ant B after changing the coordinates of the centre are represented in Figures~\ref{fig:cornerCDM_offset} and~\ref{fig:cornersfdm_offset}, respectively.

\newpage
\subsection{Removing the innermost bin}
\begin{figure}[ht]
   \centering
   \includegraphics[width=\columnwidth]{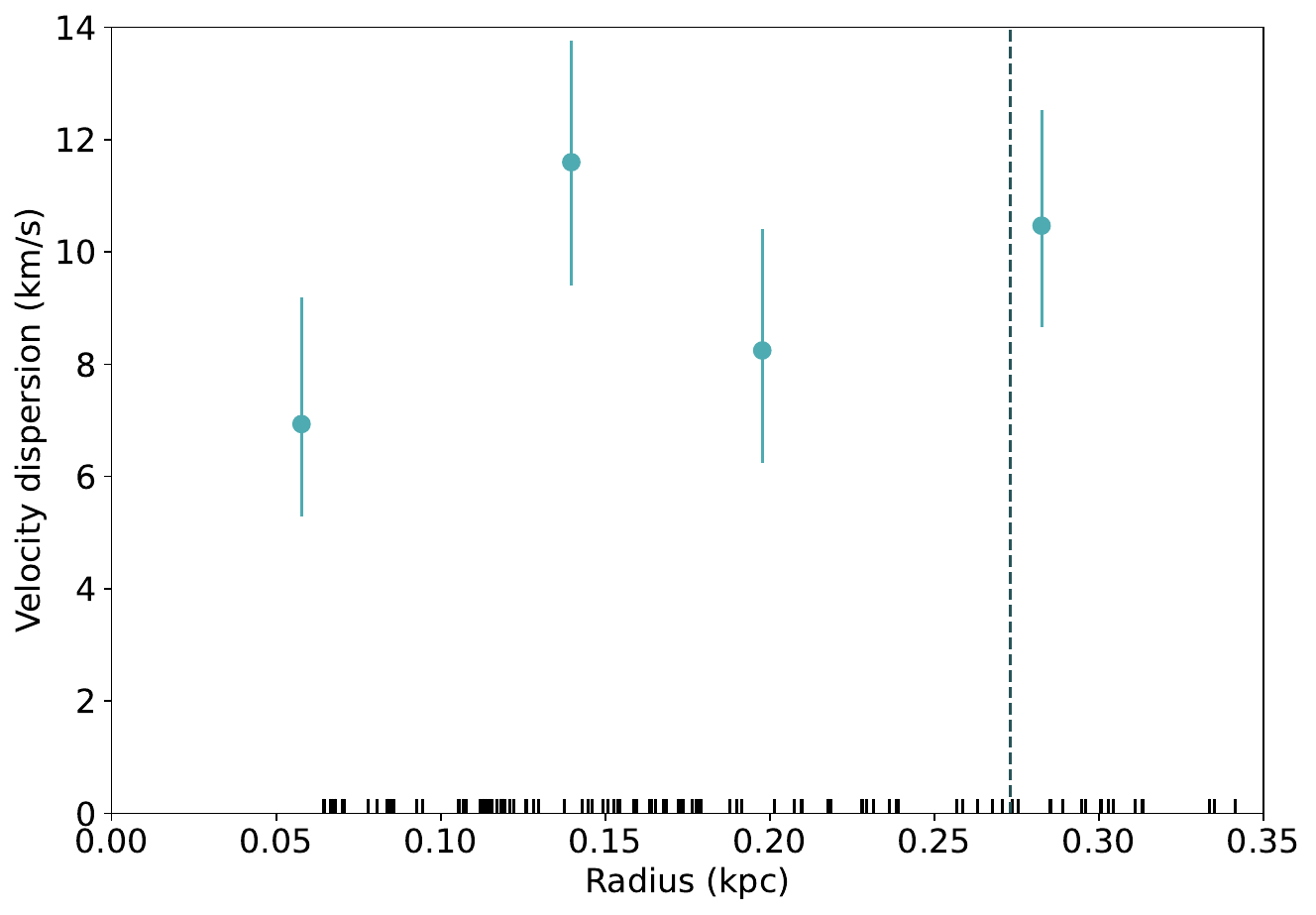}
   \caption{Velocity dispersion of each bin along with associated uncertainties for the 102 stars remaining after the removal of the 25 closest to the centre. The vertical dashed line indicates the half-light radius and the bottom marks represent the projected radii of the members of Ant B.}\label{fig:veldisp_bin_removed}
\end{figure}

\begin{figure}[ht]
   \centering
    \includegraphics[width=\columnwidth]{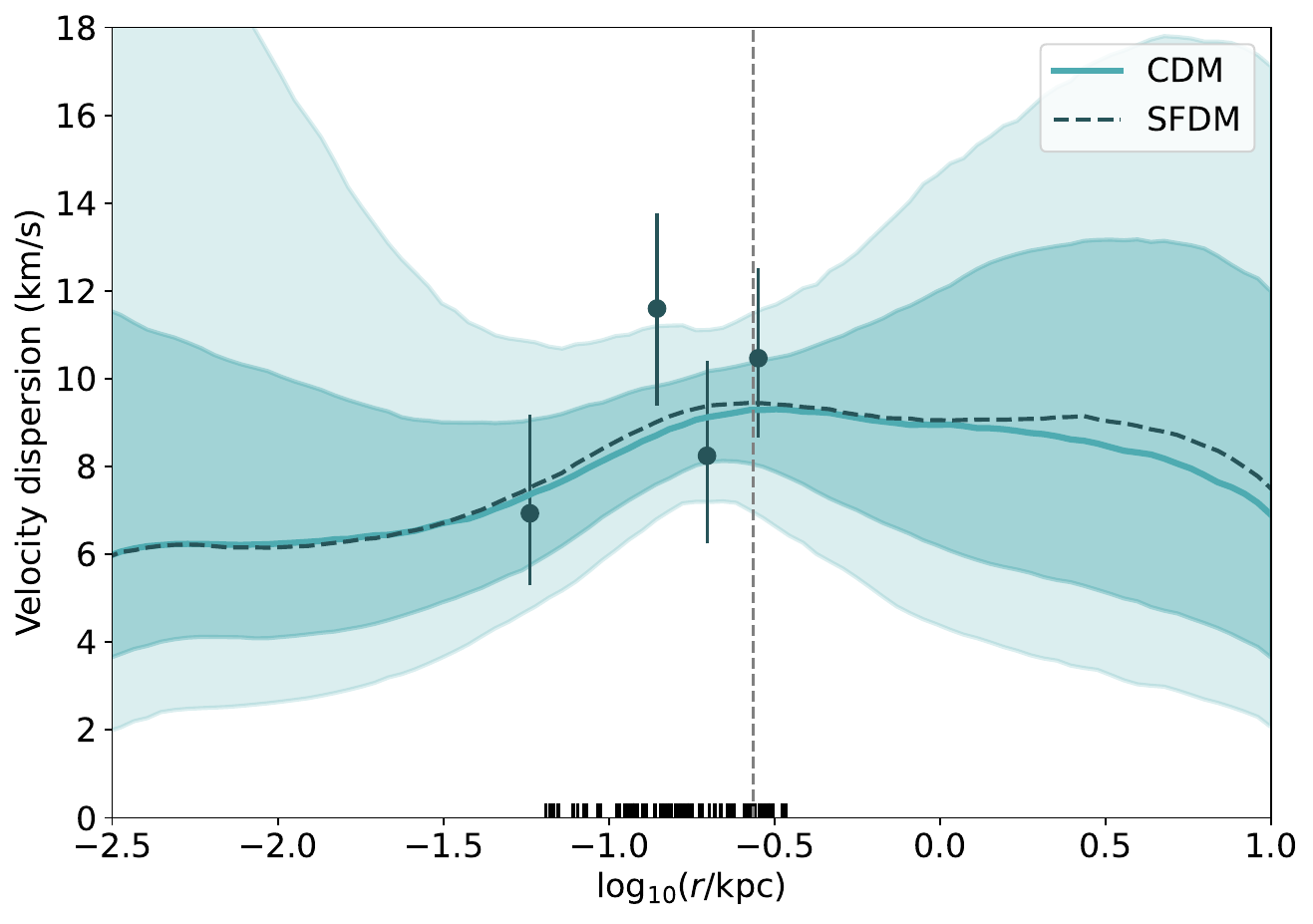}
   \caption{Velocity dispersion of Ant B without the 25 stars closest to the centre. The dark blue points represent the binned velocity dispersion previously shown. The best fit from \textsc{GravSphere} for the CDM model is shown as a solid light blue line, with the light blue shaded regions showing the 68$\%$ and 95$\%$ confidence intervals for the CDM model. The dashed dark blue line represents the best fit for the SFDM model. Since the best fit for both models is so similar, the confidence intervals for the SFDM model are omitted for clarity. The half-light radius is represented by the vertical dashed line and the bottom marks represent the projected radii of the members of Ant B.
  }\label{fig:veldisp_removed}
\end{figure}

\begin{figure}[ht]
   \centering
\includegraphics[width=\columnwidth]{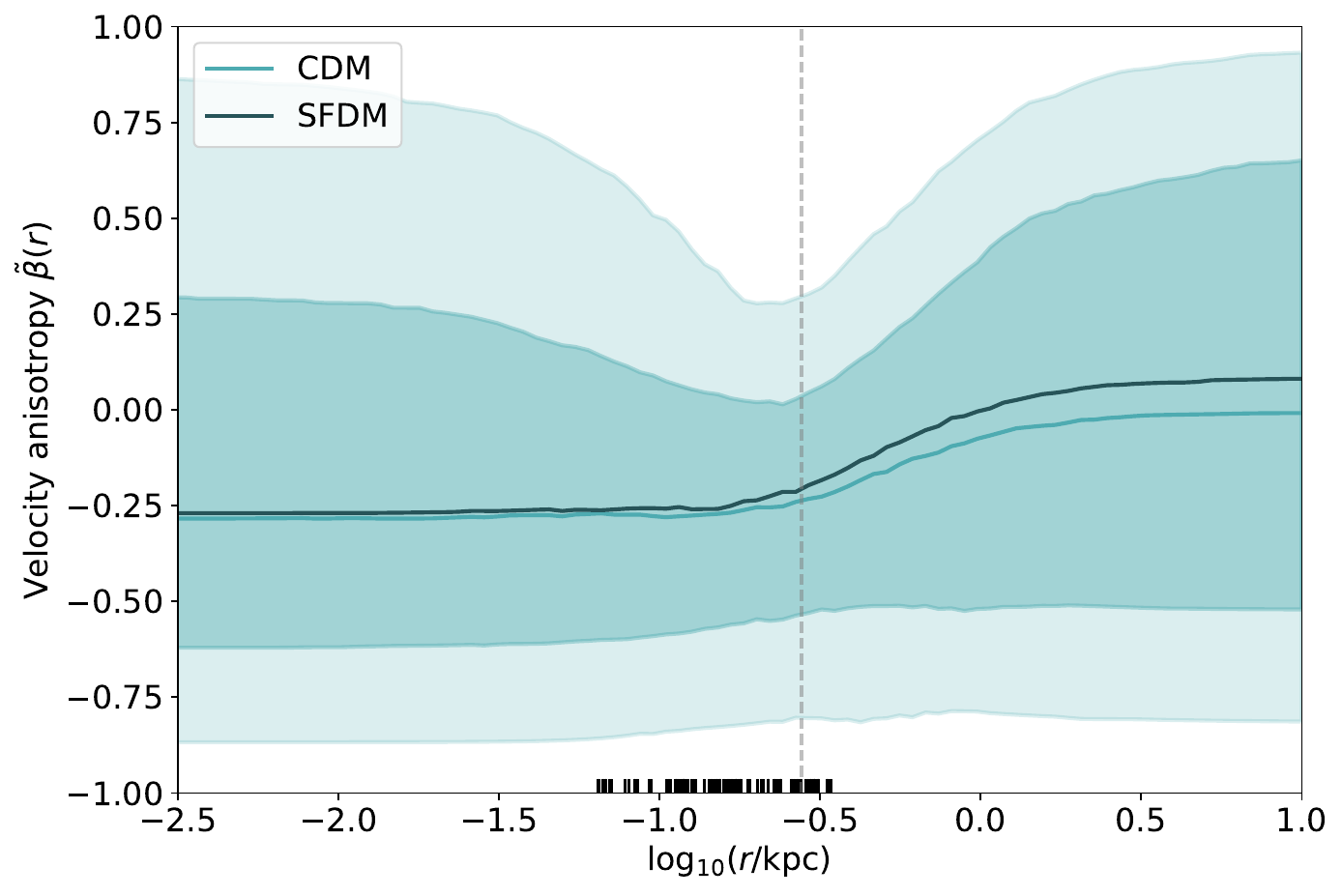}
   \caption{
   Velocity anisotropy profile of Ant B without the 25 stars closest to the centre.
   The best fit from \textsc{GravSphere} of the symmetrised anisotropy, $\Tilde{\beta}$, for the CDM model is shown as a solid light blue line, with the light blue shaded regions showing the 68$\%$ and 95$\%$ confidence intervals for the CDM model. The solid dark blue line represents the best fit for the SFDM model. Since the best fit for both models is so similar, the confidence intervals for the SFDM model are omitted for clarity. The half-light radius is represented by the vertical dashed line and the bottom marks represent the projected radii of the members of Ant B.}\label{fig:anis_removed}
\end{figure}

\begin{figure}[ht]
   \centering 
   \includegraphics[width=\columnwidth]{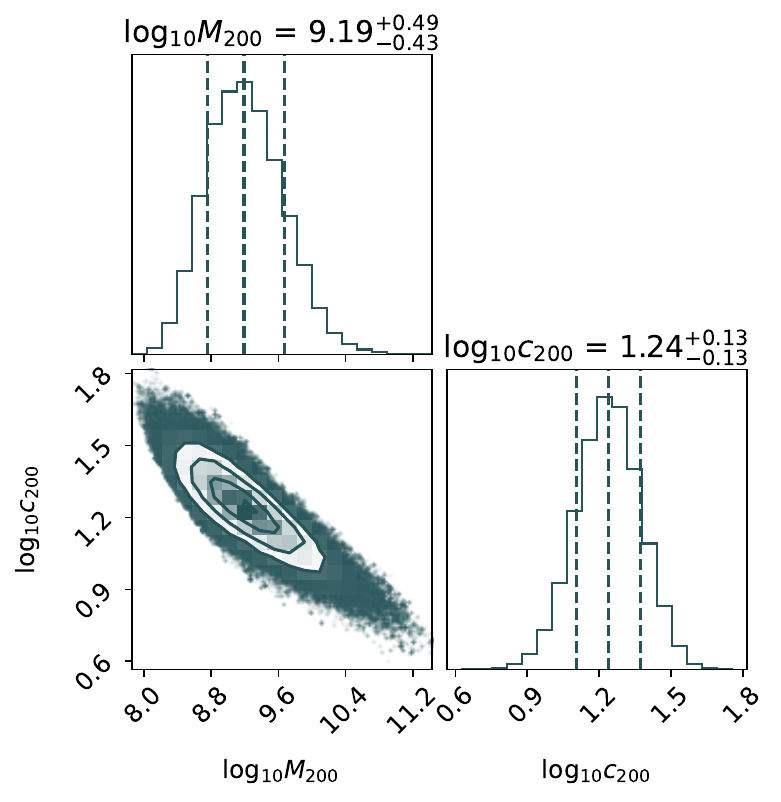}
   \caption{Constraints on the CDM profile for Ant B after removing the 25 stars closest to the centre. The histograms along the diagonal represent the posterior distribution for each parameter: the virial mass, $M_{200}$ in M$_\odot$, and the concentration parameter, $c_{200}$.}\label{fig:cornerCDM_1stbin}
\end{figure}

\begin{figure}[ht]
   \centering
   \includegraphics[width=\columnwidth]{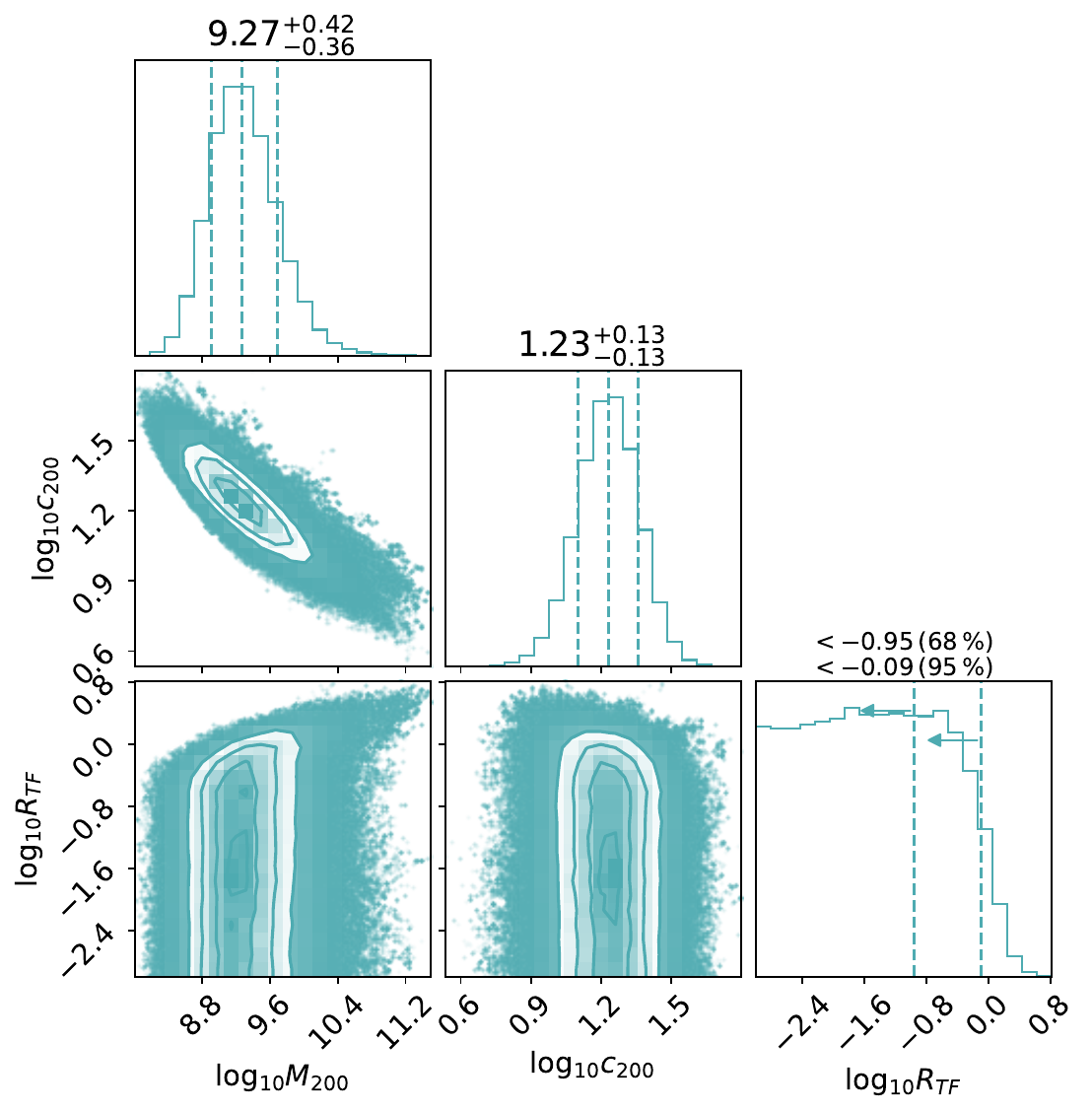}
   \caption[]{Constraints on the SFDM profile for Ant B after removing the 25 stars closest to the centre. The histograms along the diagonal represent the posterior distribution for each parameter: the virial mass, $M_{200}$ in M$_\odot$, the concentration parameter, $c_{200}$, and the characteristic length scale of the repulsive SI, $R_{\text{TF}}$ in kpc.}\label{fig:cornersfdm_1stbin}
\end{figure}

\begin{figure}
   \centering
    \includegraphics[width=\columnwidth]{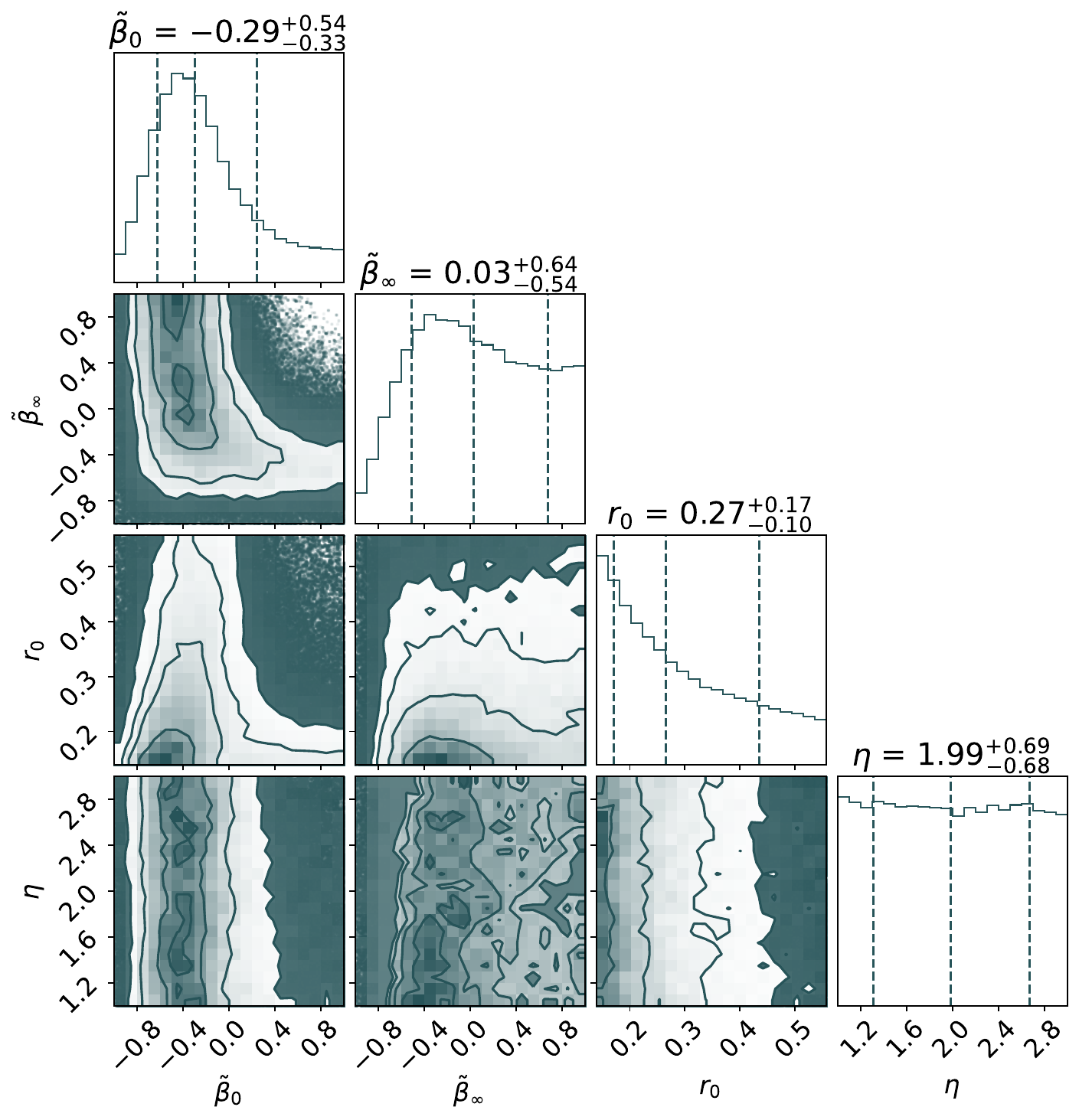} \\ 
    \includegraphics[width=\columnwidth]{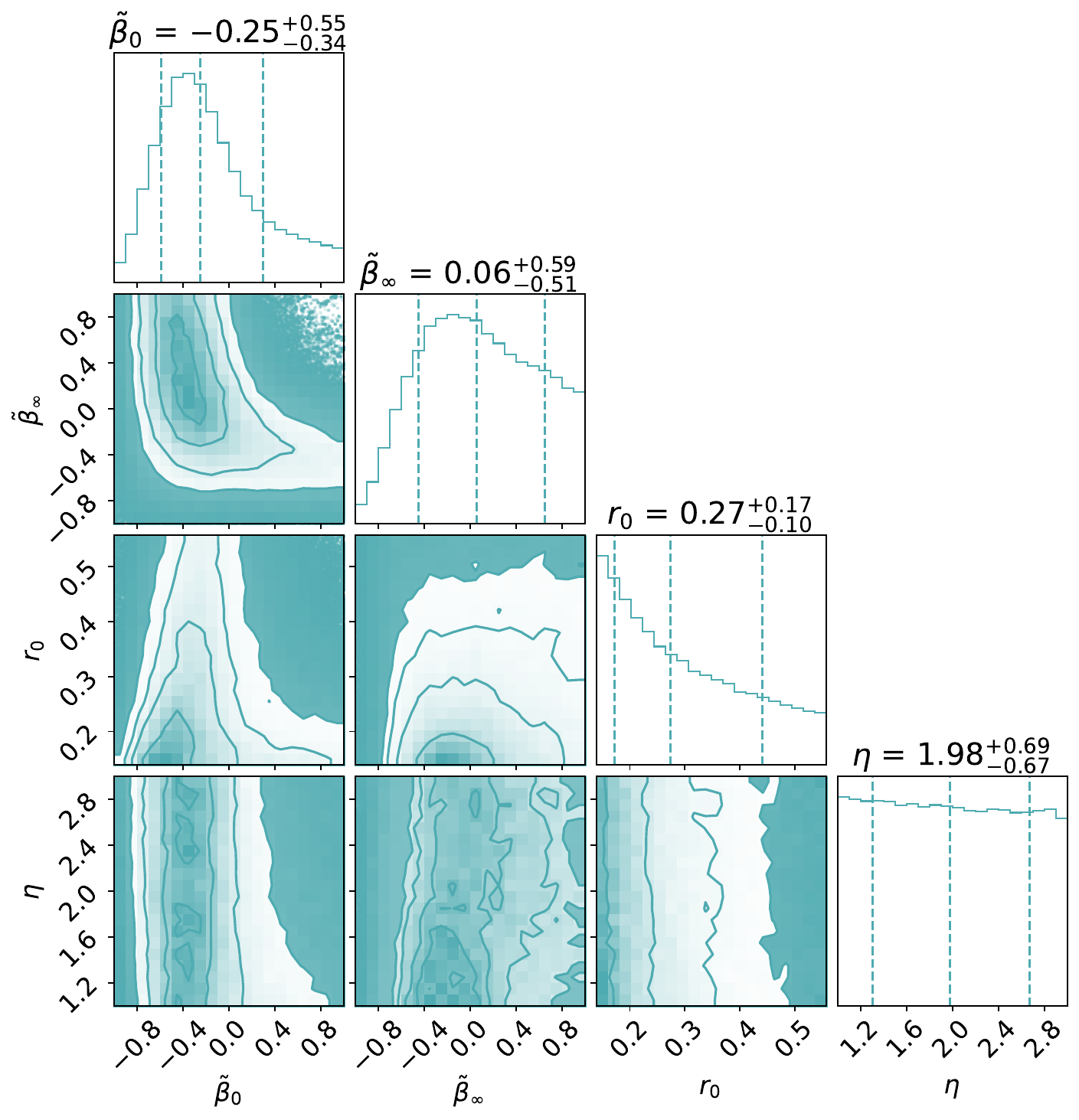}
    \caption{Corner plot of the parameters $\Tilde{\beta}_0$, $\Tilde{\beta}_\infty$, $r_0$, and $\eta$ of the anisotropy profile without the 25 innermost stars for both models. \textbf{Top:} For the CDM model. \textbf{Bottom:} For the SFDM model.}%
    \label{fig:corner_beta_removed}
\end{figure}

The second robustness test was motivated by the tangential anisotropy profile found in Ant B. The stars belonging to the first bin were removed since the first bin seemed to be the cause of this behaviour, as can be seen in the data points in Figure 4. The velocity dispersion of each bin after, without the innermost bin, the respective recovered velocity dispersion profile, and the associated anisotropy profile can be seen in Figures~\ref{fig:veldisp_bin_removed},~\ref{fig:veldisp_removed}, and~\ref{fig:anis_removed}, respectively. The constraints obtained for both the CDM and SFDM profiles for Ant B after removing the innermost bin are represented in Figures~\ref{fig:cornerCDM_1stbin} and~\ref{fig:cornersfdm_1stbin}, respectively. The posterior distributions obtained for the anisotropy parameters using this configuration can be seen in Figure~\ref{fig:corner_beta_removed}.

\end{appendix}
\end{document}